\documentclass[%
 aip,
 amsmath,amssymb,
 reprint,%
]{revtex4-1}

\usepackage{graphicx}% Include figure files
\usepackage{dcolumn}% Align table columns on decimal point
\usepackage{bm}% bold math
\usepackage[utf8]{inputenc}
\usepackage[T1]{fontenc}
\usepackage{mathptmx}
\usepackage{etoolbox}
\usepackage{subcaption} % For subfigure support
\usepackage{float}
\usepackage{hyperref}

\hypersetup{
    colorlinks=true,    % Enables colored links
    linkcolor=blue,     % Color of internal links
    citecolor=magenta,      % Color of citations
    pdfborder={0 0 0}   % Removes the red boxes around links
}

\makeatletter
\def\@email#1#2{%
 \endgroup
 \patchcmd{\titleblock@produce}
  {\frontmatter@RRAPformat}
  {\frontmatter@RRAPformat{\produce@RRAP{*#1\href{mailto:#2}{#2}}}\frontmatter@RRAPformat}
  {}{}
}%
\makeatother
\begin{document}

\preprint{AIP/123-QED}

\title{ A low-cost four-component relativistic coupled cluster linear response theory based on perturbation sensitive natural spinors}
% Force line breaks with \\
\author{Sudipta Chakraborty}
\author{Amrita Manna}
\affiliation{ 
Department of Chemistry, Indian Institute of Technology Bombay, Mumbai 400076, India%\\This line break forced with \textbackslash\textbackslash
}%

\author{T. Daniel Crawford}
\affiliation{ 
Department of Chemistry, Virginia Tech, Blacksburg, Virginia 24061, USA%\\This line break forced with \textbackslash\textbackslash
}%

\author{Achintya Kumar Dutta}

\affiliation{
Department of Chemistry, Indian Institute of Technology Bombay, Mumbai 400076, India%\\This line break forced% with \\
}%
 \email{achintya@chem.iitb.ac.in}
\affiliation{
Department of Inorganic Chemistry, Faculty of Natural Sciences, Comenius University Bratislava
Ilkovičova 6, Mlynská dolina 842 15 Bratislava, Slovakia \\
}%
\email{achintya.kumar.dutta@uniba.sk}
\date{\today}% It is always \today, today,

\begin{abstract}
We present an efficient implementation of four-component linear response coupled cluster singles and doubles (4c-LRCCSD) theory that enables accurate and computationally efficient calculation of polarizabilities for systems containing heavy elements. We have observed that the frozen natural spinor (FNS)-based truncation scheme is not suitable for linear response properties, as it leads to larger errors in static and dynamic polarizability values. In this work, we have introduced a "perturbation-sensitive" density to construct the natural spinor basis, termed FNS++. Using FNS++, we achieve excellent accuracy when compared to experimental data and other theoretical results, even after truncating nearly 70\% of the total virtual spinors. We also present pilot applications of 4c-LRCCSD method to calculate the polarizability spectra of $3d$ transition metals. By employing the FNS++-based 4c-LRCCSD, we have been able to compute polarizabilities for systems with over 1200 virtual spinors, maintaining low computational cost and excellent accuracy. 
\end{abstract}
\maketitle

\section{\label{sec:level1}Introduction:\protect\\}

Response properties, such as polarizabilities, are a fundamental phenomenon of atoms and molecules that describes their ability to redistribute electron density in response to an external electric field. This redistribution leads to the formation of induced dipoles, which influence the system's interaction with the field. Polarizability is crucial for understanding a wide range of physical phenomena and material properties, such as intermolecular interactions, chemical reactivity, and optical properties.  Fundamental properties of atomic and molecular systems, such as stiffness, compressibility, hardness, softness, and hyper-softness, are frequently associated with electric polarizability.\cite{Maroulis} Moreover, the advancement of atomic and molecular physics, as well as the development of precision technologies that depend on this response property, demands accurate calculation of polarizabilities. For example, precise values of the atomic dipole polarizability are necessary for calculating Stark and black-body radiation shift in atomic clocks, which are essential for accurate timekeeping.\cite{Sahoo21, Dong18} Accurate calculation of polarizability, one of the most fundamental linear response properties, necessitates the use of high-level theoretical approaches and well-optimized basis sets to ensure a reliable representation of electron density redistribution under external electric fields. Over the past few decades, significant progress has been achieved in the development of \textit{ab initio} methods for computing response properties, enhancing both accuracy and applicability.\cite{datta1995coupled, Hendrik77, Helgaker1990,sekino1984linear, casida1995time,christiansen1995second, christiansen1998response, christiansen1999frequency, hald2003calculation, kobayashi1994calculation,nielsen1980transition, hammond2008coupled} Most of this progress took place within the non-relativistic framework, but some efforts have been reported in the incorporation of relativistic effects for linear response properties. Relativistic effects play a crucial role in accurately describing the electronic structure of heavy atoms and molecules. The inclusion of relativistic effects in wave function-based methods is essential for achieving reliable predictions of polarizability, particularly for systems containing heavy elements. It should be noted that the inclusion of the relativistic effect is not only crucial for achieving an accurate quantitative description of systems containing heavy elements but also important for systems composed solely of light elements.\cite{pyykko1988relativistic} In addition to relativistic effects, electron correlation significantly influences the accuracy of computed polarizability values.\cite{hammond2009accurate} Relativistic polarizability calculations at the four-component (4c) Dirac-Hartree-Fock (DHF) and density functional theory (DFT) levels have been reported by Saue \textit{et al.} \cite{saue2003linear} and Salek \textit{et al.} \cite{salek2005linear}, respectively. The accuracy of relativistic DFT calculations is highly sensitive to the choice of exchange-correlation functionals, with no systematic approach available for improving these results. \cite{hait2018accurate, burke2012perspective} In contrast, wavefunction-based electron correlation methods provide greater accuracy in polarizability calculations, offering a more reliable framework for precise predictions with a scope for systematic improvement.\cite{salek2005comparison,gauss1998triple,larsen1999polarizabilities} 

The use of the four component(4c) Dirac-Coulomb (DC) Hamiltonian is considered one the most practical ways to incorporate relativistic effects in the quantum mechanical calculation of many-electron systems. The ground state reference wave function is typically constructed by performing a four-component Dirac Fock (4c-DHF)\cite{swirles1935relativistic} calculation. However, any implementation based on 4c-DHF comes with significantly higher computational costs in comparison to a standard non-relativistic method. The reason behind this additional computational overhead is spin-orbit coupling (SOC), which appears naturally in a relativistic framework due to the inherent presence of spin in the Dirac equation.\cite{dirac1928quantum,dirac1928quantumPartII,dirac1930theory} This SOC mixes electron spin and orbital angular momenta, breaking the spin and spatial symmetries, thereby necessitating the use of complex number algebra, which is generally not available in the existing non-relativistic implementations. Therefore,  electron correlation methods based on four-component Dirac-Hartree-Fock (4c-DHF) cannot use algorithms that rely on nonrelativistic spin symmetry. Instead, they must employ frameworks that incorporate time-reversal symmetry and double-group representation .\cite{dyall2007introduction} Coupled-cluster (CC) theory\cite{vcivzek1966correlation,vcivzek1969use, vcivzek1991origins,paldus2005beginnings,crawford2007introduction}  is widely regarded as one of the most precise and reliable methods among the available post-Hartree-Fock electron correlation methods for systems where a single-reference determinant predominantly describes the electronic structure. Its exponential parametrization of the wave function facilitates an appropriate description of electron correlation, guarantees size extensivity, and allows for systematic improvement. The coupled-cluster singles and doubles (CCSD) approach balances accuracy and computational cost, making it widely used for small to moderate-sized molecular systems. However, the CCSD method has a formal computational scaling of $O(N^6)$, where $N$ is the size of the correlation space. Relativistic four- and two-component Hamiltonian-based coupled cluster methods are available in literature for ground and excited-state energies,\cite{eliav1994relativistic, eliav1994open, visscher1995kramers, visscher1996formulation, lee1998spin, nataraj2010general, visscher2001formulation, koulias2019relativistic, liu2021relativistic} as well as analytic first- or higher-order property calculations.\cite{shee2016analytic, liu2021analytic, zheng2022geometry, yuan2023frequency, yuan2024formulation} 
Recent advancements in second-order static and dynamic polarizability calculation include the implementation of density-fitting-based CCSD linear response functions using the spin-free exact two-component one-electron (SFX2C1e) Hamiltonian by Dutta and co-workers,\cite{chakraborty2024spin} as well as the development by Gomes and co-workers \cite{yuan2024formulation} employing the exact two-component (X2C) Hamiltonian.  Until now, most of the second-order property calculations reported using the 4c-DC CCSD framework have been based on the finite-field approach. However, this method entails a higher computational cost, greater sensitivity to numerical noise, and inherent limitations in its extension to dynamic or frequency-dependent (time-dependent) properties. In addition, the static and dynamic polarizability values of heavy elements are extremely sensitive to the basis set used and require at least a quadruple-zeta quality basis set with sufficient numbers of diffuse functions\cite{chakraborty2024spin} However, the use of large basis sets can be extremely costly in relativistic coupled cluster calculations, as the relativistic CCSD method is approximately 32 times more expensive than the corresponding non-relativistic version.\cite{dyall2007introduction} Recently, the use of natural spinors\cite{chamoli2022reduced, surjuse2022low, yuan2022assessing, majee2024reduced}   has emerged as an attractive option to reduce the computational cost of relativistic coupled cluster calculations. Gomes and co-workers\cite{yuan2023frequency} have shown that the standard MP2-based frozen natural spinors (FNSs) do not yield systematic trends with respect to the truncation level for excitation energies and other properties. The trends are consistent with the non-relativistic framework,\cite{kumar2017frozen} where it has been observed that to accurately describe a property corresponding to an external perturbation, the natural orbital must be aware of the wavefunction's response to external perturbations\cite{d2020pno++,crawford2019reduced}. 
In this work, we present the theory and implementation of 4c-DC coupled cluster linear response theory (4c-LRCCSD) calculating the frequency-dependent polarizability of heavy elements and its low-cost implementation using perturbation-sensitive natural spinors.

\section{Theory}
\subsection{Relativistic linear response coupled cluster method}

The relativistic CC theory generates the exact wave function by employing an exponential parametrization on the reference state wave function. 
\begin{equation}
\left| {{\Psi }}_{{CC}} \right\rangle ={{e}^{{\hat{T}}}}\left| {{{\Phi }}_{0}} \right\rangle
\end{equation}
where,\begin{equation}
\hat{T}={{\hat{T}}_{1}}+{{\hat{T}}_{2}}+{{\hat{T}}_{3}}+....+{{\hat{T}}_{n}}
\end{equation}
is the cluster operator and $\left|{\Phi }_{0}\right\rangle$ is the reference Slater determinant, usually a DHF wave function. In second-quantized form the $\hat{T_{1}}$ and $\hat{T_{2}}$ operators can be expressed as
\begin{align}
\hat{T}_1 &= \frac{1}{2} \sum\limits_{ia} t_i^a \{ a_a^\dag a_i \} \\
\hat{T}_2 &= \frac{1}{4} \sum\limits_{ijab} t_{ij}^{ab} \{ a_a^\dag a_b^\dag a_j a_i \}
\end{align}
and for any \textit{n}-tuple excitation, the cluster operator can be expressed as
\begin{equation}
{\hat T_n} = {\left( {\frac{1}{{n!}}} \right)^2}\sum\limits_{ij...ab...}^n {t_{ij...}^{ab...} \{ \hat a_a^\dag \hat a_b^\dag...\;{{\hat a}_j}{{\hat a}_i}...\}} 
\end{equation}
Here, $t_{ij...}^{ab...}$ are the cluster amplitudes, $\hat{a}^\dag $, $\hat{a}$ are second quantization creation and annihilation operator, and $(i,j,k...)$ and $(a,b,c,...)$ represent occupied and virtual spinor indices respectively. 
The Schrödinger equation in terms of the coupled cluster similarity transformed Hamiltonian $(\bar{H}_{DC})$ can be written as
\begin{equation}
    {\bar{H}}_{DC}\left| {{\Phi }_{0}} \right\rangle =E\left| {{\Phi }_{0}} \right\rangle 
\end{equation}
where, ${\bar{H}_{DC}}={e}^{-\hat{T}}{\hat{H}_{DC}}{e}^{\hat{T}}$ and $\hat{H}_{DC}$ is the relativistic DC Hamiltonian, defined as
\begin{equation}
    {{\hat{H}}_{DC}}=\sum\limits_{i}^{n}{\left[ c{{{\vec{\alpha }}}_{i}}.{{{\vec{p}}}_{i}}+{{\beta }_{i}}{{m}_{0}}{{c}^{2}}+{{V}_{nuc}}({{r}_{i}}) \right]+\sum\limits_{i>j}^{n}{\frac{1}{{{r}_{ij}}}I_{4}}}
\end{equation}
where $\vec{p}_{i}$ and $m_0$  are the momentum vector and rest mass of the electron, respectively, $c$ is the speed of light, and ${V}_{nuc} = \sum\limits_{A}^{nuc}V_{iA}$, where $V_{iA}$ represents the potential energy associated with the interaction of the $i$-th electron with the electrostatic field of nucleus $A$. $\alpha$ and $\beta$ are the Dirac matrices, and $I_4$ denotes a $4 \times 4$ identity matrix. The many-body ground state reference wave function $\left| {{\Phi }_{0}} \right\rangle$ is generally obtained by solving the DHF equation, which can be expressed in matrix form as
\begin{equation}
\left[ \begin{aligned}
  & \hat{V}+\hat{J}-\hat{K}\quad \ \ \ c(\sigma .\hat{P})-\hat{K} \\ 
 & c(\sigma .\hat{P})-\hat{K}\ \ \ \hat{V}-2m_{0}{{c}^{2}}+\hat{J}-\hat{K} \\ 
\end{aligned}\right]
\left[ \begin{aligned}
  & {{\Phi }^{L}} \\ 
 & {{\Phi }^{S}} \\ 
\end{aligned} 
\right]=E
\left[ \begin{aligned}
  & {{\Phi }^{L}} \\ 
 & {{\Phi }^{S}} \\ 
\end{aligned} \right]
\end{equation}
where ${\Phi }^{L}$ and ${\Phi }^{S}$ are the large and small components of the four-component wave function, respectively, each of them having a two-spinor form. The $\hat{V}$ denotes the nucleus-electron interaction, $\hat{J}$ and $\hat{K}$ represent the Coulomb and exchange operators respectively, and $\sigma$ are the Pauli spin matrices. The electron correlation problem in the relativistic framework is generally solved using the no-pair approximation\cite{dyall2007introduction}. The relativistic CCSD energy and cluster amplitudes are obtained by projecting onto the reference and excited determinants, respectively.
\begin{align}
    {{E}_{CC}}=\left\langle  {{\Phi }_{0}} \right|\bar{H}\left| {{\Phi }_{0}} \right\rangle \\
    \left\langle  {{\mu }_{i}} \right|\bar{H}\left| {{\Phi }_{0}} \right\rangle = 0, \quad i=1,2
\end{align}
Here, $\mu_i$ refers to the singly and doubly excited determinants. Response theory can be formulated based on the coupled cluster framework for property calculation. It focuses on evaluating molecular properties that arise from the interaction of the ground state wave function with an external perturbation. Following time-dependent perturbation theory, the effect of the external field can be added to the unperturbed Hamiltonian as 
\begin{align}
    \hat{H} &= \hat{H}_0 +\hat{V(t)} \\
   \hat{ V(t)} &= \int_{-\infty}^{\infty} V(\omega) e^{(\alpha-i\omega)t}\, d\omega
\end{align}
where $\hat{H}_0$ is the unperturbed DC Hamiltonian and $\hat{V}(t)$ denotes the external perturbation or the interaction operator, which vanishes at $t=-\infty$. The $V(\omega)$ is the Fourier transform of $V(t)$ and $\alpha$ represents a real positive infinitesimal quantity, such that $V(-\infty) = 0$. When an external perturbation is applied, the wave function acquires a time-dependent nature and evolves dynamically following the time-dependent Schrödinger equation,
\begin{equation}
    i\frac{d}{dt} \left| {\Psi }_{0} (t)\right\rangle = (\hat{H}_0 + \hat{V}(t))\left| {\Psi }_{0}(t)\right\rangle
\end{equation}
and the time-dependent wave function $\left| {\Psi }_{0} (t)\right\rangle$ can be expressed as,\cite{olsen1985linear}
\begin{equation}
    \left| {\Psi }_{0} (t)\right\rangle = \left| {\overline{\Psi} }_{0}\right\rangle e^{i\epsilon(t)}
\end{equation}
where $\epsilon(t)$ is a real phase factor, and $\left| {\overline{\Psi} }_{0}\right\rangle$ is the phase-independent part of the wave function that can be expanded in terms of perturbation order as,
\begin{equation}
   \left| {\overline{\Psi} }_{0}\right\rangle = \left| {\Psi }_{0}\right\rangle + {\left| {\Psi }_{0}\right\rangle}^{1} + {\left| {\Psi }_{0}\right\rangle}^{2} + ...
\end{equation}
Now to examine the time evolution of the expectation value of an operator $\hat{A}$, we can express it in terms of the time-dependent wave function $(\left| {\Psi }_{0} (t)\right\rangle)$ as,
\begin{equation}
  \left\langle \hat{A} \right\rangle (t)  = \left\langle  {{\Psi }_{0} (t)} \right|\hat{A}\left| {{\Psi }_{0} (t)} \right\rangle
\end{equation}
and it can also be expanded as,\cite{olsen1985linear}
\begin{eqnarray}
\left\langle \hat{A} \right\rangle(t)=&&\left\langle  {{\Psi }_{0}} \right|\hat{A}\left| {{\Psi }_{0}} \right\rangle + \int_{-\infty}^{\infty}{\left\langle \left\langle \hat{A}; V^{\omega_1}\right\rangle \right\rangle}_{\omega_1 + i\alpha} e^{-i(\omega_1 + i\alpha)} \, d\omega_1 \nonumber \\ 
&&+ \frac{1}{2}\int_{-\infty}^{\infty} d\omega_1 \int_{-\infty}^{\infty} d\omega_2 {\left\langle \left\langle \hat{ A}; V^{\omega_1}; V^{\omega_2}\right\rangle \right\rangle}_{\omega_1 + i\alpha,\, \omega_2 + i\alpha} \nonumber \\
&&\times e^{-i(\omega_1 + \omega_2 + 2i\alpha)} + ...
\end{eqnarray}
where $ {\left\langle \left\langle \hat{A}; V^{\omega_1}\right\rangle \right\rangle}_{\omega_1 + i\alpha} $, and ${\left\langle \left\langle \hat{A}; V^{\omega_1}; V^{\omega_2}\right\rangle \right\rangle}_{\omega_1 + i\alpha,\, \omega_2 + i\alpha}$ denote the linear and quadratic response functions, respectively. From Eq.(17), it can also be seen that the linear response function represents the first-order perturbational contribution to the expectation value of the time-independent operator $\hat{A}$ and the linear response function for exact states can be expressed as,
\begin{eqnarray}
{\left\langle \left\langle \hat{A}; V^{\omega_1}\right\rangle \right\rangle} =&&\sum\limits_{k}\left[\frac{\left\langle  {{\Psi }_{0}} \right|\hat{A}\left| {{\Psi }_{k}} \right\rangle\left\langle  {{\Psi }_{k}} \right|V^{\omega_1}\left| {{\Psi }_{0}} \right\rangle}{\omega_1 - \omega_k}\right] \nonumber \\
&&-\sum\limits_{k} \left[ \frac{\left\langle  {{\Psi }_{0}} \right|V^{\omega_1}\left| {{\Psi }_{k}} \right\rangle\left\langle  {{\Psi }_{k}} \right|\hat{A}\left| {{\Psi }_{0}} \right\rangle}{\omega_1 + \omega_k} \right]
\end{eqnarray}
where $\omega_k$ is the excitation energy corresponding to the transition from the ground state $({\Psi }_{0})$ to the $k$-th excited state $({\Psi }_{k})$. The summation runs over all the excited states, and Eq.(18) is also known as the sum-over-states equation. The calculation of linear response properties using this equation is practically not feasible for larger systems since it involves the computation of all the excited states.

The computationally demanding sum-over-states approach can be circumvented if the CCSD wave function is parameterized in a suitable manner. The recipe to calculate coupled cluster response property in a non-relativistic framework has been presented by Koch et al.\cite{koch1990coupled} and time-dependence of the right $(\left| {\Psi }_{cc} (t)\right\rangle)$ and left $(\left\langle {\Lambda } (t)\right|)$ CCSD wave functions can be expressed as,
\begin{gather}
        \left| {\Psi }_{cc} (t)\right\rangle = e^{\hat{T}(t)}\left| {\Phi }_{0} (t)\right\rangle e^{i\epsilon(t)} \\
    \left\langle {\Lambda } (t)\right| = \{ \left\langle  {{\Phi }_{0}} \right| + \sum\limits_{\mu} \lambda_\mu (t) \left\langle  {\mu } \right| e^{-\hat{T}(t)} \} e^{-i\epsilon(t)}
\end{gather}
where, $e^{\pm i\epsilon(t)}$ is the time-dependent phase factor. From Eq.(13), one can obtain the expression for time derivatives of cluster amplitudes ( $t_\mu$ and $\lambda_\mu$), which can be written as,
\begin{align}
    \frac{dt_\mu}{dt} = -i \left\langle  {\mu}\right|e^{-{\hat{T}(t)}}(\hat{H}_0 + \hat{V}(t))e^{\hat{T}(t)}\left| {{\Phi }_{0}} \right\rangle \\
    \frac{d\lambda_\mu}{dt} = i \left\langle  {\tilde{\Lambda}(t)}\right|\left[\hat{H}_0 + \hat{V}(t), \tau_\mu\right]\left| {\tilde{{\Psi }}_{cc}(t)} \right\rangle
\end{align}
where,
\begin{gather}
        \left\langle  {\widetilde{\Lambda}(t)}\right|= \left\langle  {{\Phi }_{0}} \right| + \sum\limits_{\mu} \lambda_\mu (t) \left\langle  {\mu } \right| e^{-\hat{T}(t)} \\
    \left| {\widetilde{\Psi }}_{cc} (t)\right\rangle = e^{\hat{T}(t)}\left| {\Phi }_{0} (t)\right\rangle
\end{gather}

We can expand the time-dependent cluster amplitudes in terms of perturbation order as
\begin{align}
    t_\mu(t) &= t_{\mu}^{(0)} + t_{\mu}^{(1)}+t_{\mu}^{(2)}+...\\
    \lambda_\mu(t) &= \lambda_{\mu}^{(0)} + \lambda_{\mu}^{(1)} + \lambda_{\mu}^{(2)}+...
\end{align}
and application of Fourier transformation on Eq.(25) and (26) will lead to,
\begin{align}
    t_\mu(t) &= t_{\mu}^{(0)} + \int_{-\infty}^{\infty}X_{\mu}^{(1)} e^{(-i\omega_1 + \alpha)} \, d\omega_1 + ...\\
    \lambda_\mu(t) &= \lambda_{\mu}^{(0)} + \int_{-\infty}^{\infty}Y_{\mu}^{(1)} e^{(-i\omega_1 + \alpha)} \, d\omega_1 +...
\end{align}
where $X_{\mu}^{(1)}$ and $Y_{\mu}^{(1)}$ are the Fourier transforms of $t_{\mu}^{(1)}$ and $\lambda_{\mu}^{(1)}$,  respectively. The expression for solving $X_{\mu}^{(1)}$ is 
\begin{align}
    X_{\mu}^{(1)}(\omega_1+i\alpha) &= \sum_{\nu} \left[-\mathbf{A}+(\omega_1 + i\alpha)\mathbf{I}\right]^{-1}_{\mu\nu}\xi_{\nu}^{(1)}(\omega_1)
\end{align} 
where $\mathbf{I}$ is an identity matrix and $\mathbf{A}$ denotes the Coupled Cluster Jacobian, expressed as
\begin{equation}
    A_{\mu\nu} = \left\langle  {\mu} \right| \left[ \bar{H}_0,\tau_\nu\right] \left| {{\Phi }_{0}} \right\rangle
\end{equation}
and
\begin{equation}
    \xi_{\nu}^{(1)}(\omega_1) = \left\langle  {\nu} \right|\overline{V}{^{(\omega_1)}}\left| {{\Phi }_{0}} \right\rangle
\end{equation}
Similarly for $Y_{\mu}^{(1)}$,
\begin{align}
  && Y_{\mu }^{(1)}({{\omega }_{1}}+i\alpha )\text{ }=-\sum\limits_{\nu }{\{\eta _{\nu }^{(1)}}({{\omega }_{1}})\text{ } \nonumber \\ 
 && +\sum\limits_{\gamma }{{{F}_{\nu \gamma }}}X_{\gamma }^{(1)}({{\omega }_{1}}+i\alpha )\} \nonumber \\ 
 && \times \left\{ \mathbf{A}+({{\omega }_{1}}+i\alpha )\mathbf{I} \right\}_{\nu \gamma }^{-1}  
\end{align}

where, $\eta^{(1)}$ and matrix $\mathbf{F}$ are defined as
\begin{align}
    &\eta_{\nu}^{(1)}(\omega_1) = \left\langle  {(1 + \hat{\Lambda})} \right|[\,\overline{V}{^{(\omega_1)},\tau_\nu}]\left| {{\Phi }_{0}} \right\rangle, \\
    &F_{\nu\gamma} = \left\langle  {(1 + \hat{\Lambda})} \right|[[\bar{H}_{0}, \tau_\nu],\tau_\gamma ]]\left| {{\Phi }_{0}} \right\rangle .
\end{align}
Now the time-dependent expectation value of the CC wave function can also be expanded in terms of orders of perturbation as shown in Eq.(17) and the linear response function within the CC framework can be defined as,
\begin{align}
    \left\langle \left\langle \textbf{A};\textbf{B} \right\rangle \right\rangle &= \frac{1}{2} \hat{P}(A,B) \left[ \left\langle \Phi_0 \left| [Y_{\omega_1}^{B}, \bar{A}] \right| \Phi_0 \right\rangle \right. \notag \\
    &\quad + \left. \left\langle \Phi_0 \left|(1+\hat{\Lambda}) [\bar{A},X_{\omega_1}^{B}] \right| \Phi_0 \right\rangle \right]
\end{align}
Here, $B=V^{\omega_1}$ and the operator $\hat{P}(A, B)$ simultaneously swaps the positions of operators $\hat{A}$ and $\hat{B}$ and applies complex conjugation to the resulting expression. $X_{\omega_1}^{B}$ and $Y_{\omega_1}^{B}$ are the perturbed right and left-hand Coupled Cluster amplitudes, respectively, for the operator $\hat{B}$. An alternative approach of solving the CC response function circumvents the computation of $Y_{\omega_1}^{B}$ amplitudes by instead solving an additional set of $X_{-\omega_1}^{B}$ amplitudes as,
\begin{equation}
\begin{aligned}
\langle\langle \mathbf{A}; \mathbf{B} \rangle\rangle_{\omega_{1}} &=  
\frac{1}{2} \hat{C}^{\pm \omega_{1}} \hat{P} [A(-\omega_{1}), B(+\omega_{1})] 
\Big\langle \Phi_0 \Big| (1 + \hat{\Lambda}) \\
&\quad \times \Big( [\bar{A}, \hat{X}^{B}_{\omega_{1}}] 
+ \frac{1}{2} [[\bar{H}, \hat{X}^{A}_{-\omega_{1}}], \hat{X}^{B}_{\omega_{1}}] \Big) 
\Big| \Phi_0 \Big\rangle
\end{aligned}
\label{eq:correlation_function}
\end{equation}
Here \( \hat{C} \) is a symmetrization operator that simultaneously reverses the sign of the external field frequency and applies the complex conjugate to the entire expression, while \( \hat{P} \) ensures symmetrization of the operators $\hat{A}$ and  $\hat{B}$. Both the approaches defined in Eq.(35) and (36) are computationally equally expensive, and our current implementation of 4c-LRCCSD is based on Eq.(35) where we solve both $X_{\omega_1}^{B}$ and $Y_{\omega_1}^{B}$ amplitudes.

\subsection{Natural spinor}
The natural orbitals are defined as the eigenfunctions of the correlated one-body reduced density matrix.\cite{lowdin1955quantum} Similarly, one can obtain natural spinors by diagonalizing a spin-coupled  one-body reduced density obtained from an electron correlation method in a relativistic framework. In the frozen natural spinors (FNS) framework, the occupied spinors are kept frozen at their DHF description. The unrelaxed one-body reduced density matrix (RDM) at the MP2 level can be expressed as:
\begin{equation}
    \Gamma_{pq} = \left\langle  {\Psi^{(1)}} \right| \{ a_{p}^{\dag}a_q\} \left| {\Psi^{(1)}} \right\rangle
\end{equation}
where $\left| {{\Psi }^{(1)}} \right\rangle$ represents the first-order correction to the DHF wave function, which is defined as,
\begin{equation}
    \left| {{\Psi }^{(1)}} \right\rangle = \frac{1}{4}\sum_{ijab} t_{ij}^{ab}\left| \Phi_{ij}^{ab} \right\rangle
\end{equation}
where
\begin{equation}
    t_{ij}^{ab} = \frac{\left\langle  {ij} \right|  \left| {ab} \right\rangle}{\epsilon_i + \epsilon_j - \epsilon_a - \epsilon_b}
\end{equation}
Here, $\left\langle  {ij} \right|  \left| {ab} \right\rangle$ is anti-symmetrized two-electron integral and $\epsilon_i$, $\epsilon_a ...$ denote the DHF molecular spinor energies and $\left| \Phi_{ij}^{ab} \right\rangle$ is the doubly excited determinant. The virtual-virtual block of the relativistic MP2 one-body RDM can be expressed as,
\begin{equation}
    \Gamma_{ab} = \frac{1}{2}\sum_{ijc}t_{ij}^{ac}t_{ij}^{bc}
\end{equation}
By diagonalizing the one-body RDM,
\begin{equation}
    \boldsymbol{\Gamma} \mathbf{V} = \mathbf{\textit{n}}\mathbf{V}
\end{equation}
one can obtain the eigenvectors, $\mathbf{V}$ and the eigenvalues, $n$. The eigenvectors and eigenvalues of Eq.(41) represent the virtual natural spinors and associated occupation numbers, respectively. The  wave function exhibits significantly larger sparsity when expressed in terms of natural spinor basis compared to the original canonical molecular spinor basis. Spinors with lower occupation numbers (\(n\)) yield coupled cluster  amplitudes of smaller magnitude, contributing less to the correlation energy. As a result, natural spinors with occupation numbers below a certain threshold can be safely discarded without any significant impact on the accuracy of the calculations. Using truncated virtual natural spinors, \( \boldsymbol{\tilde{V}} \), the virtual-virtual block of the canonical Fock matrix is subsequently transformed into the basis of  truncated virtual natural spinors.
\begin{equation}
    \boldsymbol{\tilde{F}}\mathbf{_{VV}}  = \boldsymbol{\tilde{V}}\boldsymbol{^\dag} \mathbf{F_{VV}} \boldsymbol{\tilde{V}}
\end{equation}
Subsequently, a semi-canonicalization is performed by diagonalizing the $\boldsymbol{\tilde{F}}\mathbf{_{VV}}$,
\begin{equation}
    \boldsymbol{\tilde{F}}\mathbf{_{VV}} \boldsymbol{\tilde{Z}} = \boldsymbol{\tilde{Z}} \boldsymbol{\tilde{\varepsilon }}
\end{equation}
where, $\boldsymbol{\tilde{Z}}$ and $\boldsymbol{\tilde{\varepsilon }}$ represent semi-canonical virtual-virtual spinors and the corresponding orbital energies, respectively. Finally, the canonical DHF virtual spinors are transformed into the virtual natural spinor basis using a transformation matrix, $\textbf{B}$, defined as
\begin{equation}
B=\tilde{Z}\tilde{V}
\end{equation}
\subsection{Perturbation sensitive natural spinor}

Analogous to the construction of natural spinors from the ground-state MP2 density, the natural spinor basis 4c-LRCCSD method can also be formulated using a perturbation sensitive one-electron density, which we refer to as FNS++. Since the polarizability is a second-order property,  second-order perturbed densities provide a more reliable tool for constructing natural spinors for such calculations. In contrast, natural spinors derived from the ground-state MP2 density do not carry meaningful information for correlated response calculations.
In the FNS++ scheme, the virtual-virtual block of the second-order one body  reduced density for a perturbation operator $\hat{A}$, can be expressed as,
\begin{equation}
    {{[{{D}^{A}_{ab}}]}^{(2)}}=\frac{1}{2}{{\left[ t_{ij}^{ac}(A) \right]}^{(1)}}{{\left[ t_{ij}^{bc}(A) \right]}^{(1)}}+{{\left[ t_{i}^{a}(A) \right]}^{(1)}}{{\left[ t_{i}^{b}(A) \right]}^{(1)}}
\end{equation}
where,
\begin{align}
  & {{\left[ t_{ij}^{ac}(A) \right]}^{(1)}}=\frac{\bar{A}_{ij}^{ac}}{{{{\bar{H}}}_{aa}}+{{{\bar{H}}}_{cc}}-{{{\bar{H}}}_{ii}}-{{{\bar{H}}}_{jj}}+\omega}  \\ 
 & {{\left[ t_{i}^{a}(A) \right]}^{(1)}}=\frac{\bar{A}_{i}^{a}}{{{{\bar{H}}}_{aa}}-{{{\bar{H}}}_{ii}}+\omega}
\end{align}
 
and 
\begin{align}
  & \bar{A}_{ij}^{ac}=\hat{P}(a,c)t_{ij}^{ec}A_{e}^{a}-\hat{P}(i,j)t_{mj}^{ac}{{A}_{i}}^{m} \\ 
 & {{{\bar{H}}}_{ii}}={{F}_{ii}}+\frac{1}{2}t_{in}^{ef}\left\langle \left. in \right\| \right.\left. ef \right\rangle  \\ 
 & {{{\bar{H}}}_{aa}}={{F}_{aa}}-\frac{1}{2}t_{mn}^{fa}\left\langle \left. mn \right\| \right.\left. fa \right\rangle
\end{align}
By replacing the density matrix with the one defined in Eq. (45) and applying Eqs. (41)–(44) as outlined in Section II.B, an FNS++ basis can be constructed.
One can justify the equation (46) and (47), as an approximation to equation (29), if only the diagonal elements of the Jacobian matrix $\hat{A}$
are retained.\cite{crawford2019reduced}
For the case of dipole polarizability, the matrix elements of $A$ corresponding to the $x$, $y$, and $z$ direction can be different, and in the present implementation, the perturbation-sensitive density is generated by taking an average of the matrix elements for all three directions. It should be noted that the perturbation-sensitive one-particle density matrix is not positive definite; therefore, the absolute value of the occupation number must be considered for the cutoff. 
The steps involved  in FNS++ 4c-LRCCSD are as follows: \\
(a) Solve the DHF equation and construct ground state reference wave function.\\
(b) Perform a relativistic ground-state MP2 calculation to obtain the unperturbed double-excitation cluster operator. \\
(c) Solve for the perturbed first-order singles and doubles amplitudes and construct the second-order perturbed densities.(Eq.(45-48)).\\
(d) Diagonalize the virtual-virtual block of the density matrix and obtain the perturbation-sensitive natural spinors.\\
(c) Truncate the virtual space on the basis of a chosen occupation threshold. \\
(d) Perform a LRCCSD calculation on the truncated virtual space at a particular frequency ($\omega$). \\

It has been observed that a perturbative correction for the excluded virtual spinors to the 4c-CCSD correlation energy leads to an improvement in the results. The correction is defined as the difference between the 4c-MP2-level correlation energy in the full untruncated space and in the truncated FNS basis,
\begin{equation}
\Delta_{4c-MP2}= E_{4c-MP2}^{canonical}-E_{4c-MP2}^{truncated}
\end{equation}
This correction is added to the CCSD correlation energy in the truncated basis to obtain a corrected 4c-CCSD correlation energy,
\begin{equation}
E_{4c-CCSD}^{corrected}= E_{4c-CCSD}^{uncorrected} + \Delta E_{4c-MP2}
\end{equation}
By same analogy, we have also added a perturbative correction to the polarizability values in the truncated basis,
\begin{equation}
\alpha_{4c-CCSD}^{corrected}= \alpha_{4c-CCSD}^{uncorrected} + \Delta \alpha_{4c-MP2}
\end{equation}
where,
\begin{equation}
\Delta \alpha_{4c-MP2} = \alpha_{4c-MP2}^{canonical}-\alpha_{4c-MP2}^{truncated}
\end{equation}
The MP2-level polarizability calculations are performed by constructing the linear response function with the 4c-MP2 level amplitudes and perturbed ${{\left[ t_{ij}^{ac}(A) \right]}^{(1)}}$ amplitudes defined in Eq.(46). 

\section{Computational Details}
The 4c-LRCCSD  method and its low-cost  FNS++ version have been implemented in our in-house software package BAGH,\cite{dutta2023bagh}, which is primarily written in Python, and the computationally bottleneck parts have been written in Cython and Fortran. BAGH is currently interfaced with  PySCF,\cite{pyscf2020, Qiming2015, Qiming2018} GAMESS-US,\cite{Barca2020} and DIRAC.\cite{DIRAC_saue2020}
To validate the accuracy of our implementation, we first computed the dynamic polarizabilities of Zn, Cd, and Hg across a representative frequency range using the canonical basis and then extended our study to the FNS++ basis. The selection of truncation criteria for FNS++ calculations within the 4c-LRCCSD framework is thoroughly discussed, followed by a systematic investigation of basis set effects, including benchmark comparisons across various Dyall basis sets. Details regarding the truncation thresholds and the specific basis sets employed for different systems are provided in the subsequent section. All calculations presented in this work are performed using the BAGH software package with the PySCF interface.

\section{Results and Discussion}

\subsection{Polarizability of Group IIB atoms}

In a closed-shell atom’s ground state, weak external electric fields do not cause a first-order energy shift. Instead, the energy shifts arise from second-order effects, which are usually expressed in terms of atomic polarizabilities. Among these atomic polarizabilities, the dipole polarizability is the most significant. Precise experimental values are available for the static dipole and dynamic dipole polarizabilities at some selected frequencies of group-IIB atoms Zn, Cd, and Hg.\cite{goebel1995dispersion, goebel1996theoretical, goebel1996dipole, hohm2022dipole}. An accurate understanding of the polarizabilities of these atoms is valuable for various fundamental applications. For instance, Cd and Hg are being considered as candidates for atomic clocks,\cite{tyumenev2016comparing, yamaguchi2019narrow} and precise polarizability data is essential for evaluating systematic effects in such clocks. Based on this perspective, we have first tested our implementation of 4c-LRCCSD on frequency-dependent polarizabilities of Zn, Cd and Hg. Figure 1 presents the dynamic polarizabilities of group IIB elements over a frequency range of 0.0 to 0.3 a.u.\ using s-aug-dyall.v2z basis set. Within this frequency region, the polarizability spectrum of Zn and Cd consists of four poles labeled A, B, C, and D. Poles A and C correspond to spin-forbidden \({}^{1}S_{0} \to {}^{3}P_{1}\) transitions, while poles B and D correspond to spin-allowed \({}^{1}S_{0} \to {}^{1}P_{1}\) transitions. In the case of Hg, only poles near regions A and B are observed within this frequency range.  In addition to the 4c-LRCCSD results, we also compare non-relativistic LRCCSD(NR-LRCC) and spin-free exact two-component LRCCSD (SFX2C1e-LRCC) values. It is observed that, at frequency values near the excitation energies corresponding to the spin-allowed transitions ${}^{1}{{S}_{0}}\to {}^{1}{{P}_{1}}$, both the SFX2C-1e and 4c-LRCCSD methods exhibit large, prominent poles. However, the poles associated with the spin-forbidden transitions ${}^{1}{{S}_{0}}\to {}^{3}{{P}_{1}}$ are only visible in the 4c-LRCCSD method. Similar findings were reported by Yuan \textit{et al.},\cite{yuan2024formulation} using the X2C Hamiltonian based LR-CCSD method. The non-relativistic curve consistently deviates from the relativistic one, though the difference between the non-relativistic and relativistic dynamic polarizabilities of the Zn atom remains relatively small. However, this deviation becomes increasingly more pronounced as we move to the heavier elements, Cd and Hg, where the relativistic effects are far more significant. The poles originating due to spin-forbidden transitions display a clear broadening as we move from Zn to Hg, signifying the growing impact of spin-orbit coupling. This broadening is relatively modest in the Zn atom but becomes more noticeable in Cd and reaches its maximum in Hg. The significant widening of the poles in Hg highlights the dominant role of spin-orbit interaction in heavier elements. The trends are consistent with the X2C-LRCCSD results reported by Yuan \textit{et al.}\cite{yuan2024formulation} Furthermore, the precise determination of the pole positions corresponding to spin-allowed and spin-forbidden transitions validates the accuracy of our 4c-LRCCSD implementation for dynamic polarizability calculations.
\begin{figure*}
\centering
    % First subplot
    \begin{subfigure}{0.48\textwidth} % Adjust width as needed
        \includegraphics[width=\linewidth]{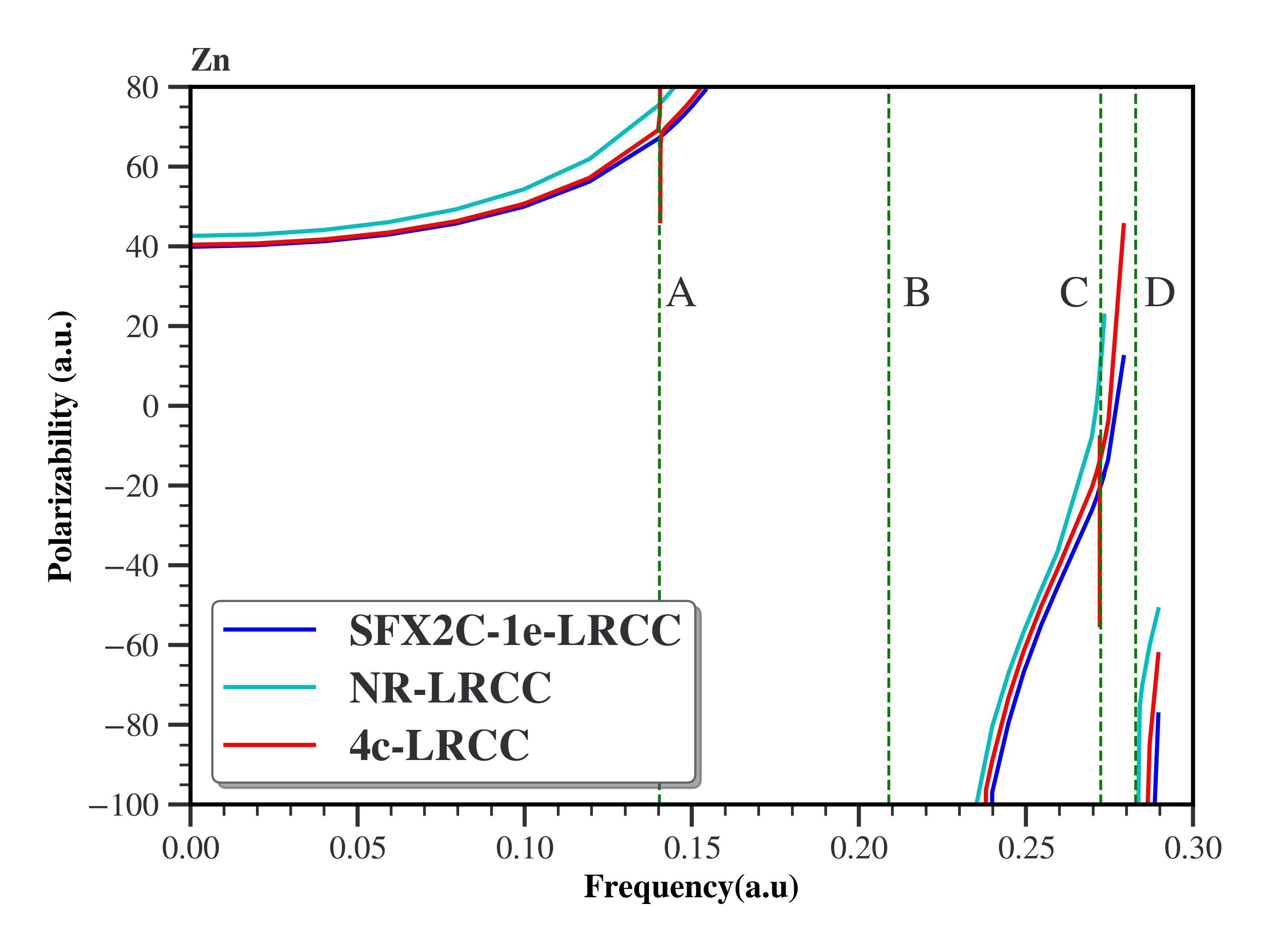} % Replace with your file
        \caption{}
        \label{fig:sub1}
    \end{subfigure}
    \begin{subfigure}{0.48\textwidth}
        \includegraphics[width=\linewidth]{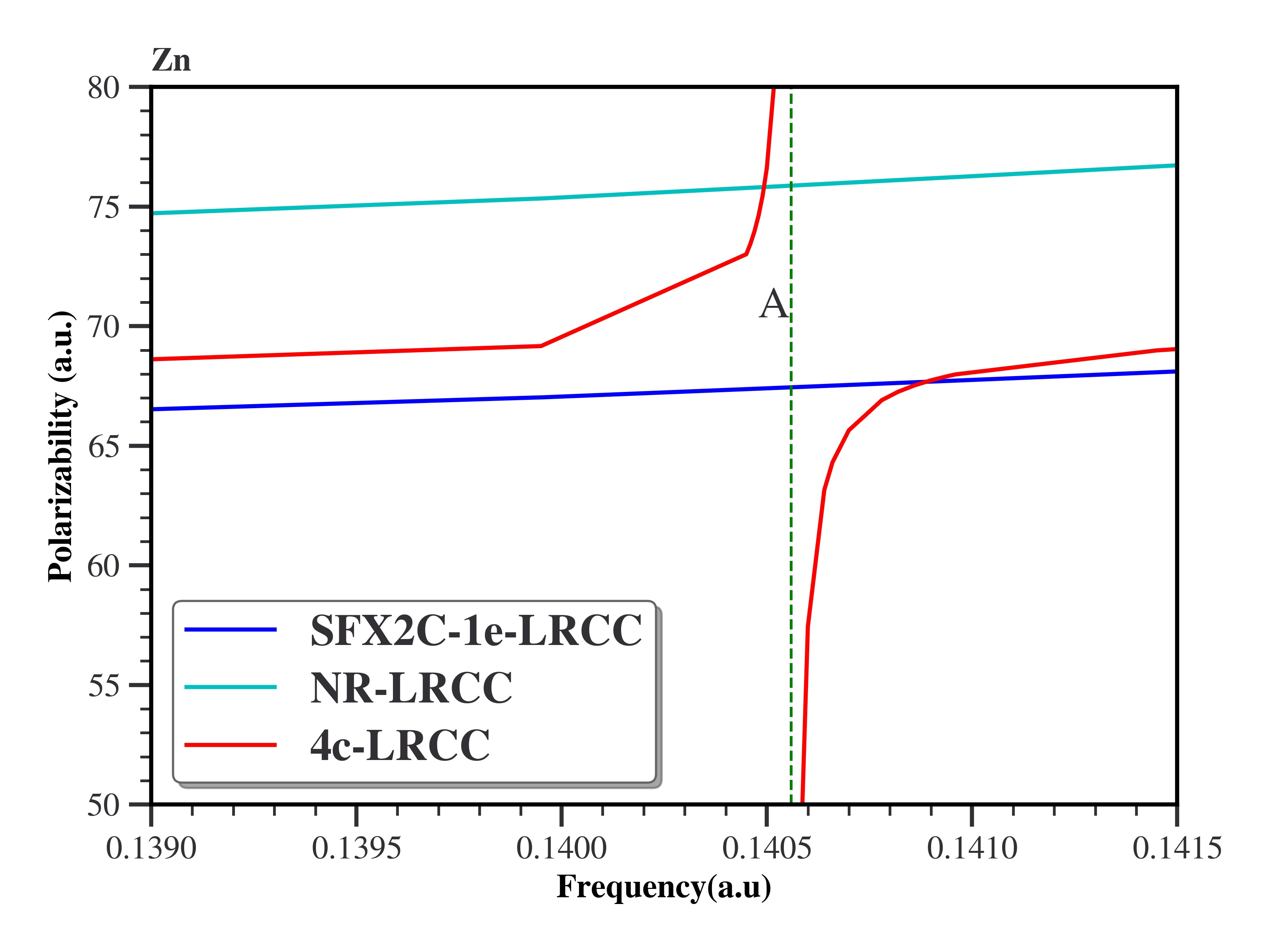}
        \caption{}
        \label{fig:sub2}
    \end{subfigure}
    
    % Second column (three figures)
    \begin{subfigure}{0.48\textwidth}
        \includegraphics[width=\linewidth]{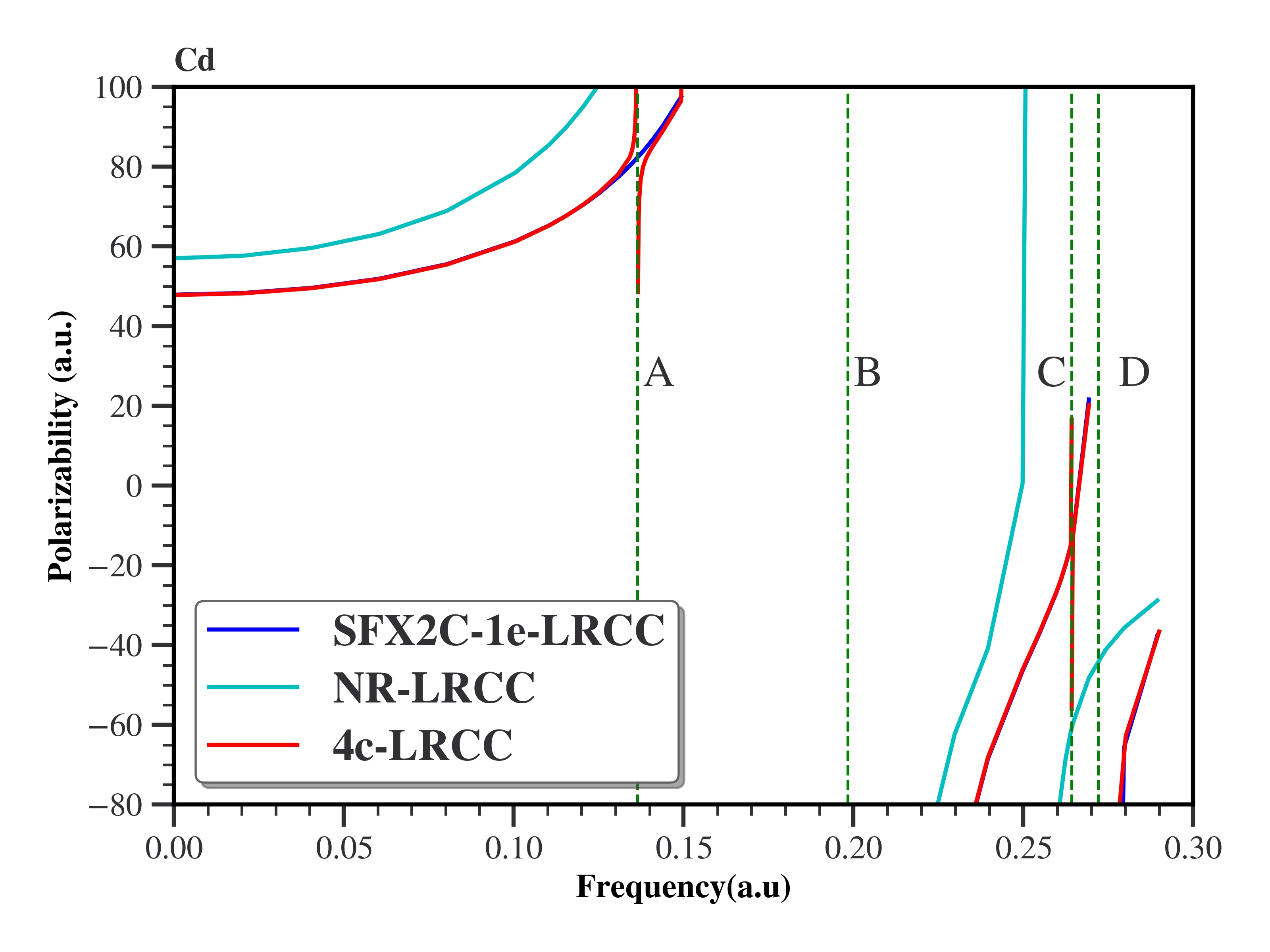}
        \caption{}
        \label{fig:sub4}
    \end{subfigure}
    \begin{subfigure}{0.48\textwidth}
        \includegraphics[width=\linewidth]{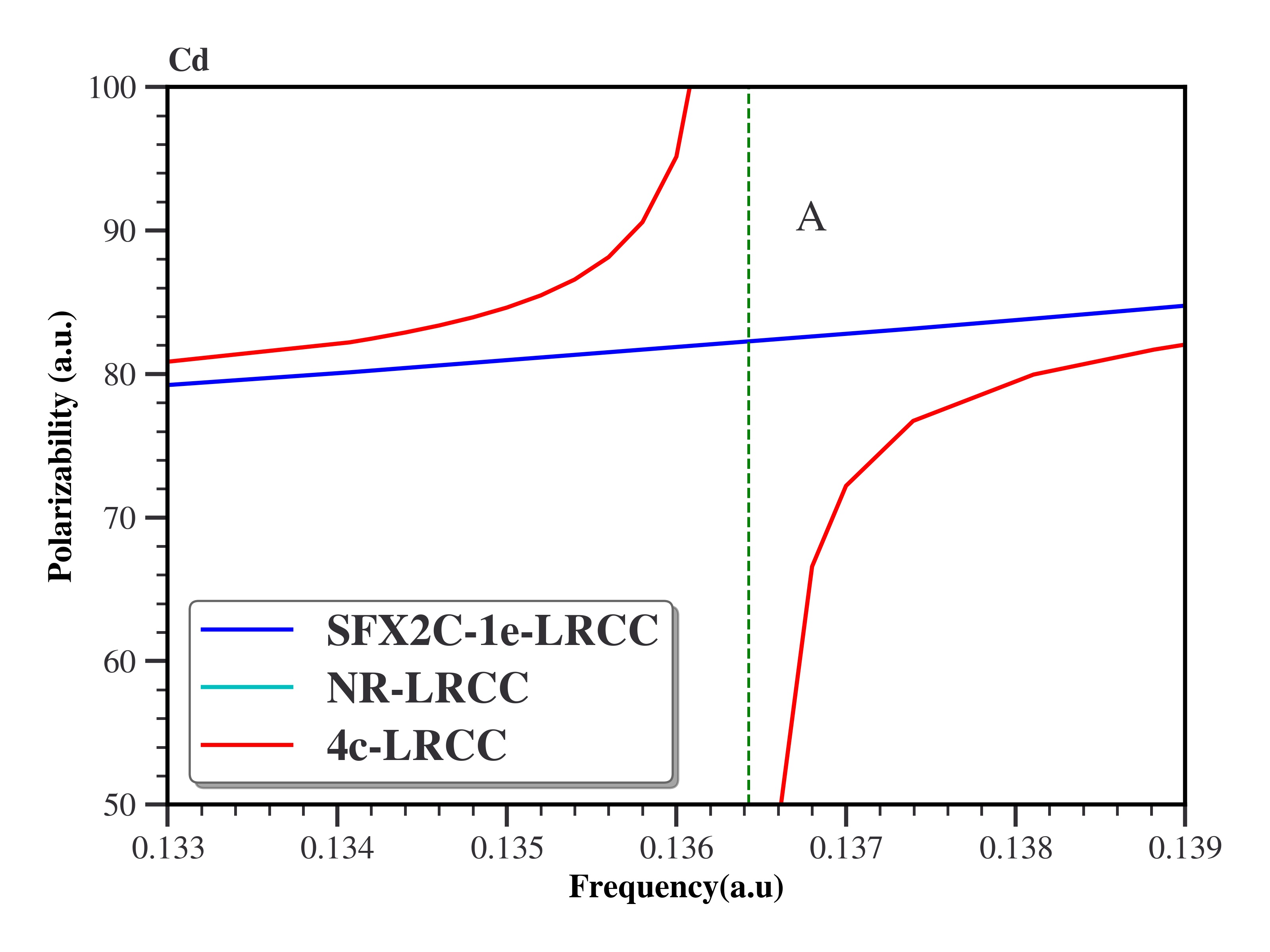}
        \caption{}
        \label{fig:sub5}
    \end{subfigure}

    % Second column (three figures)
    \begin{subfigure}{0.48\textwidth}
        \includegraphics[width=\linewidth]{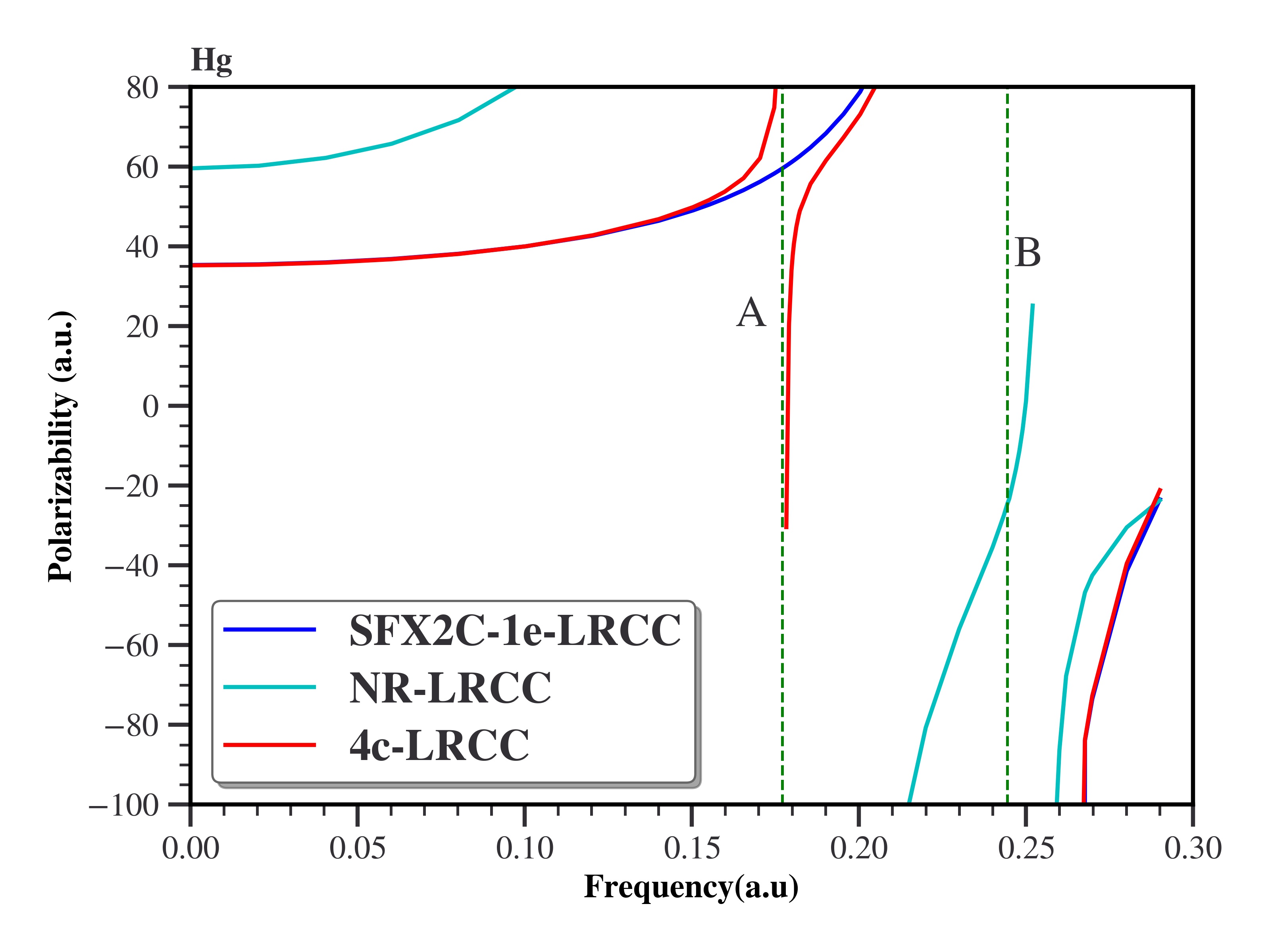}
        \caption{}
        \label{fig:sub4}
    \end{subfigure}
    \begin{subfigure}{0.48\textwidth}
        \includegraphics[width=\linewidth]{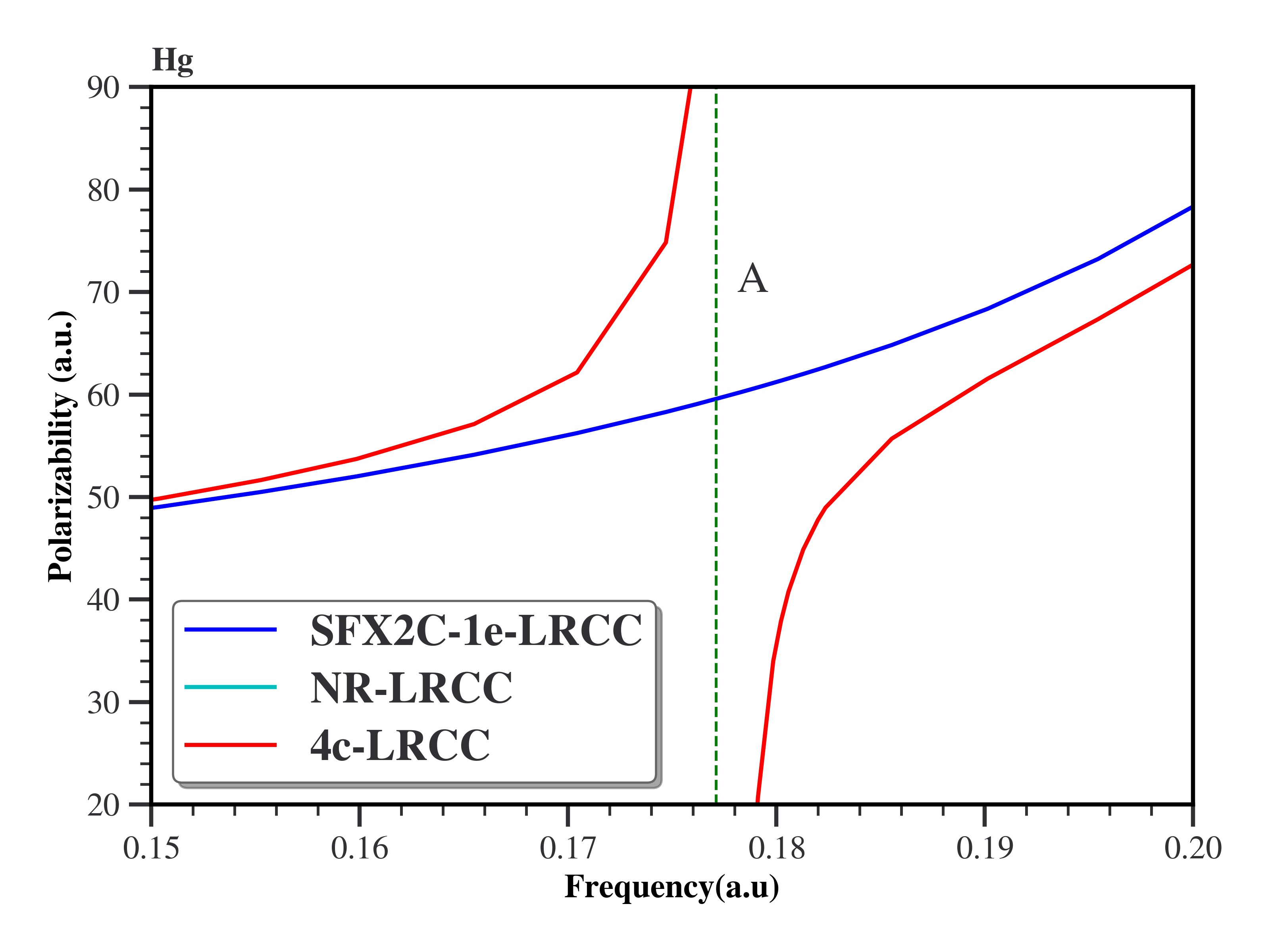}
        \caption{}
        \label{fig:sub5}
    \end{subfigure}
    % Main caption for the figure
    \caption{\label{fig:epsart}Polarizability spectra of (a) Zn, (c) Cd and (e) Hg using 4c-LRCC, SFX2C1e-LRCC and non-relativistic LRCCSD (NR-LRCC) with the s-aug-dyall.v2z basis set. The right panel depicts the zoomed-in region of the pole near the first spin-forbidden ${}^{1}{{S}_{0}}\to {}^{3}{{P}_{1}}$ transitions for (b) Zn, (d) Cd and (f) Hg}
\end{figure*}

\subsection{Comparison of FNS and FNS++ bases for static and dynamic polarizabilities}
In the relativistic CCSD framework, it has been demonstrated that truncation of the virtual orbitals in the canonical spinor basis introduces significant errors in the correlation energy.\cite{chamoli2022reduced} In contrast, discarding natural spinors with low occupation numbers allows one to recover the correlation energy with high accuracy, showcasing the efficiency of the natural spinor approach in the 4c-relativistic framework.\cite{chamoli2022reduced} However, the same is not true for response properties. Figures 2(a) and 2(b) showcase the convergence of the FNS and FNS++ basis in calculating the static polarizability and dynamic polarizability at 0.072 a.u., respectively, for the Zn atom in the uncontracted aug-cc-pVDZ basis set. The polarizability value at the untruncated canonical basis has been taken as a reference. For Zn, the static and dynamic (at 0.072 a.u. frequency) polarizability calculated using untruncated canonical basis are 39.12 a.u. and 43.52 a.u., respectively. The plot clearly demonstrates that the percentage of virtual spinors retained has a significant impact on the accuracy of the calculated polarizability values for Zn atom. One can see that the static and dynamic polarizability values converge very slowly in the FNS basis and it requires almost 90 percent of the virtual space to get close to the reference value.On the other hand, the error in the polarizability value computed in the FNS++ basis exhibits smooth convergence with respect to the size of the virtual space and shows significantly lower errors compared to the FNS method. 
Both methods eventually converge to values near the polarizability value calculated with a canonical basis as more virtual spinors are included. However, for the static polarizability, even with nearly 40\% of the virtual spinors truncated, FNS++ basis gives an impressive accuracy, with an error margin of just ~0.8\%. In contrast, considering the same scenario, the FNS basis shows a much larger error of around 32\%, highlighting its limitation in generating an accurate depiction of the wave function within a truncated basis, especially for second-order property calculation in the 4c relativistic framework. A similar trend is observed in both methods for dynamic polarizability when the external frequency is set to 0.072 a.u..

To further examine the accuracy of FNS++ for molecular systems, HBr was chosen as a test system, and the uncontracted aug-cc-pVDZ basis set was used for the calculations. The static polarizability of 22.09 a.u. obtained in an untruncated canonical basis has been used as a reference. As can be seen in Figure 2(c), the trend in the HBr result is very similar to that observed for Zn atom, where FNS performs poorly and only gives accurate polarizability values when at least 70\% of the virtual spinor space is included in the calculation. In contrast, the convergence is achieved in FNS++ basis with only 20\% of the total virtual spinor included in the correlation space. In the FNS framework, higher truncation leads to significant errors in polarizability, as the omitted spinors contribute negligibly to the energy but affect the polarizability value significantly. Conversely, FNS++ effectively selects the natural spinors, which are important for the second-order response property, allowing for accurate results with fewer number of virtual spinors. The superior performance of FNS++ in comparison to FNS with the inclusion of a very small number of virtual spinors directly translates to a reduction in computational cost and makes FNS++ an attractive choice for large-scale second-order response property calculations, where the inclusion of the entire virtual space is prohibitively expensive.
%Fig.~\ref{fig:epsart}%
\begin{figure*}
\centering
    % First subplot
    \begin{subfigure}{0.32\textwidth} % Adjust width as needed
        \includegraphics[width=\linewidth]{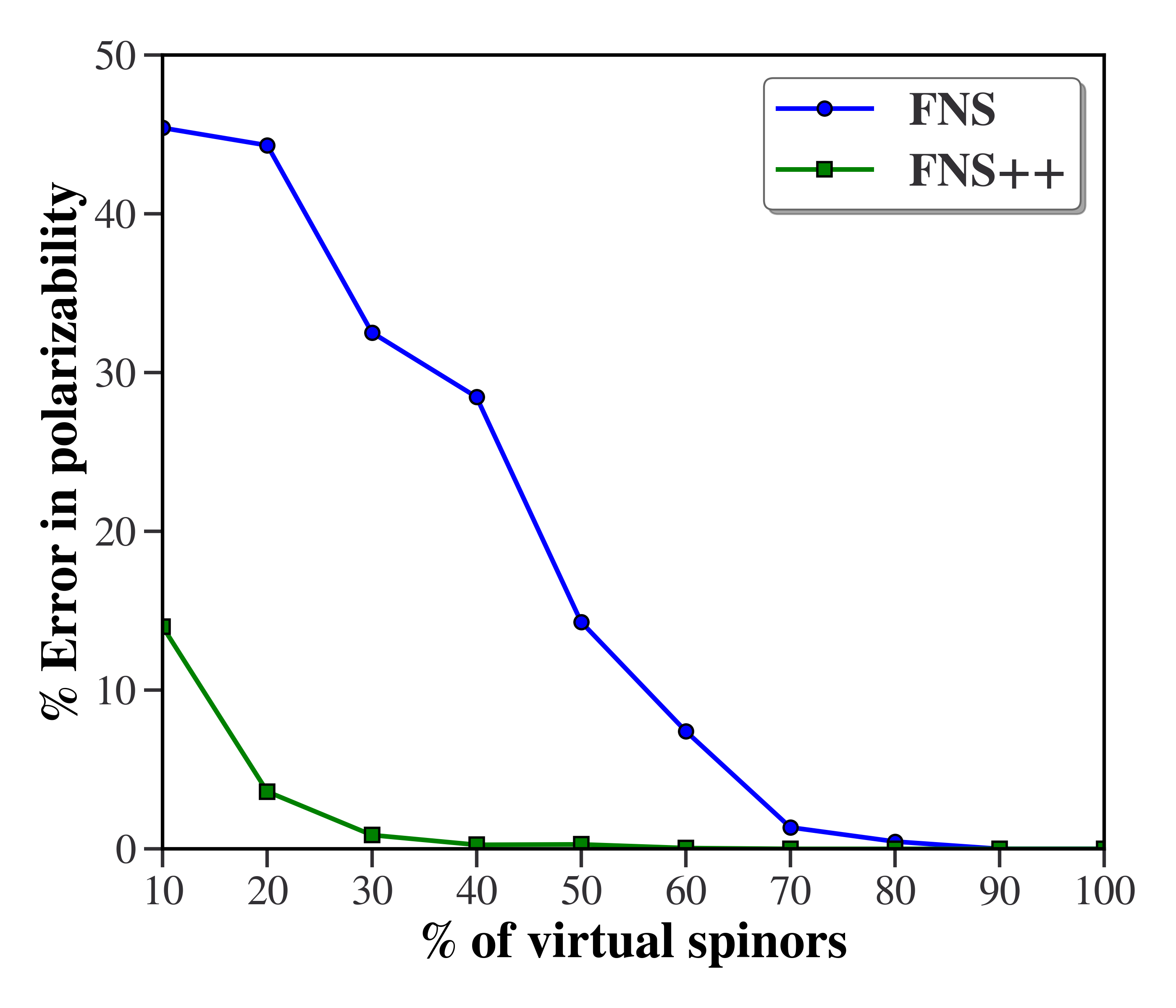} % Replace with your file
        \caption{}
        \label{fig:sub1}
    \end{subfigure}
    \begin{subfigure}{0.32\textwidth} % Adjust width as needed
        \includegraphics[width=\linewidth]{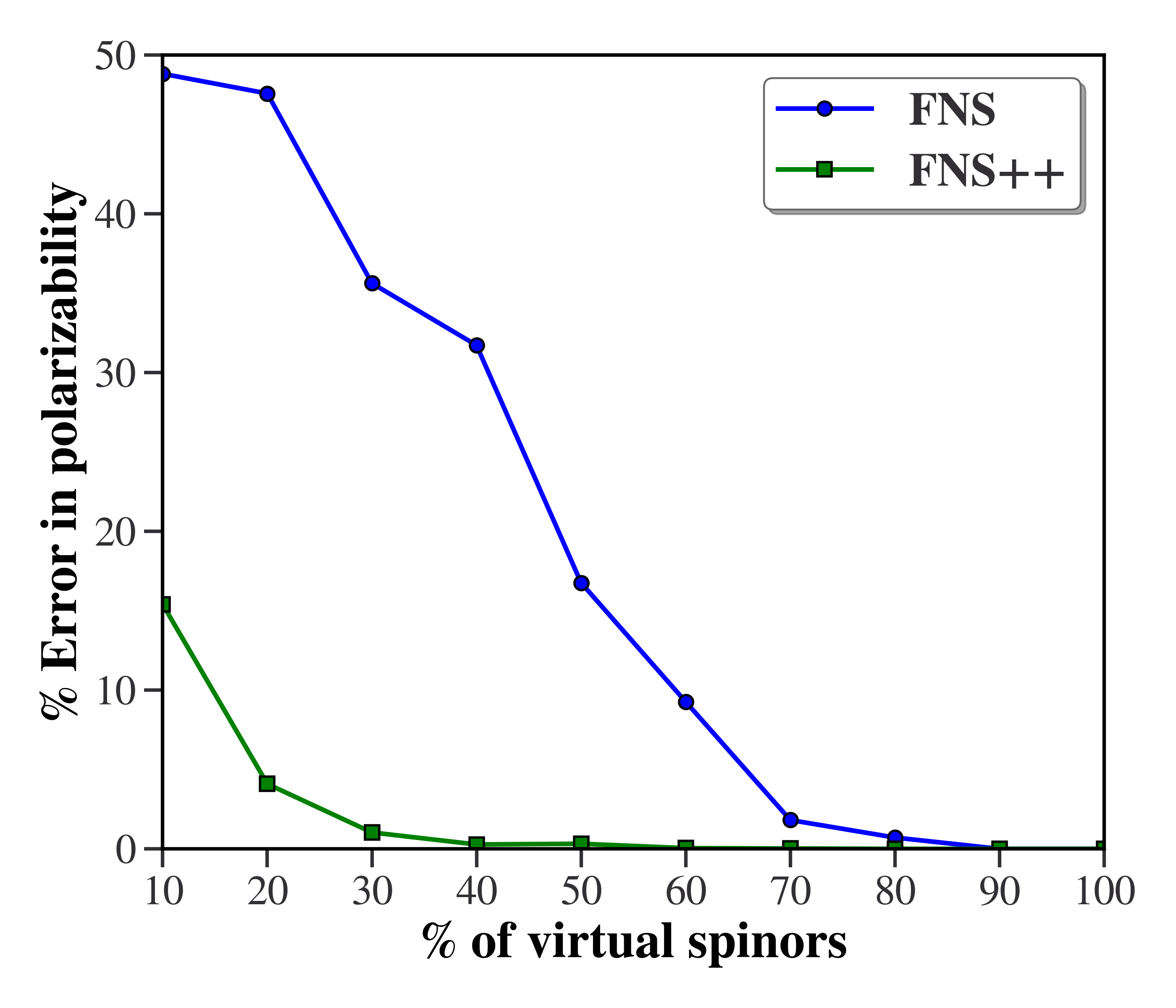} % Replace with your file
        \caption{}
        \label{fig:sub1}
    \end{subfigure}
    \begin{subfigure}{0.32\textwidth}
        \includegraphics[width=\linewidth]{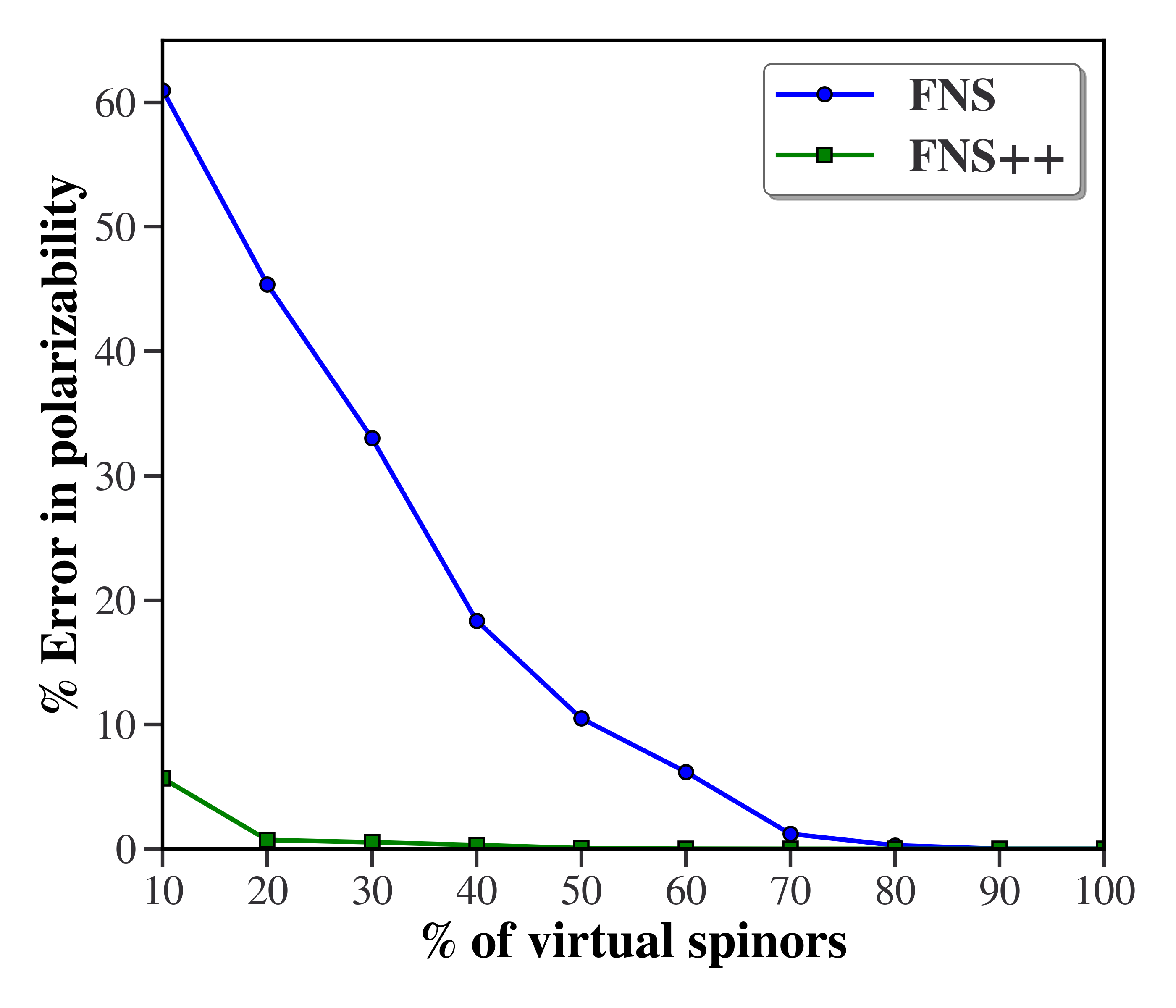}
        \caption{}
        \label{fig:sub2}
    \end{subfigure}
    
    % Main caption for the figure
    \caption{\label{fig:epsart}Error in (a) static and (b) dynamic (0.072 a.u.) polarizability of Zn  atom and (c) static polarizability of HBr with uncontracted aug-cc-pVDZ basis set in FNS and FNS++ basis.}
\end{figure*}
This analysis demonstrates that the perturbation-sensitive natural spinors, as utilized in FNS++, provide a more compact and efficient basis for polarizability calculations than the standard natural spinors in FNS basis. However, selecting natural spinors based on their occupation numbers, rather than solely on the percentage of virtual spinors retained, provides a more targeted and efficient truncation strategy.  The plot presented in Figure 3 depicts the absolute error in dynamic polarizability at 0.072 a.u.\ for the Zn atom as a function of the occupation threshold. It can be seen that, at low occupation thresholds, the calculations with the MP2-level correction result in a slightly smaller error compared to the uncorrected FNS++. Dutta and coworkers\cite{chamoli2022reduced, majee2024reduced} have shown that the inclusion of perturbative correction leads to much quicker convergence of the correlation energy with respect to the truncation threshold.   A similar scenario is observed for the polarizability calculation in the FNS++ framework.  At a low occupation threshold, such as at 10$^{-2}$, FNS++ shows less error with the perturbative correction. The FNS++ and corrected FNS++ methods show a significant reduction in the absolute error as the threshold increases, and the corrected version reaches convergence nearly at 10$^{-5}$ natural spinor occupation threshold.  The uncorrected version converges at a threshold of  \(10^{-6}\). Therefore, an occupation threshold of \(10^{-5}\) with perturbative correction is selected as the default for subsequent calculations unless explicitly stated otherwise.

\begin{figure}
\centering
    \includegraphics[width=0.5\textwidth]{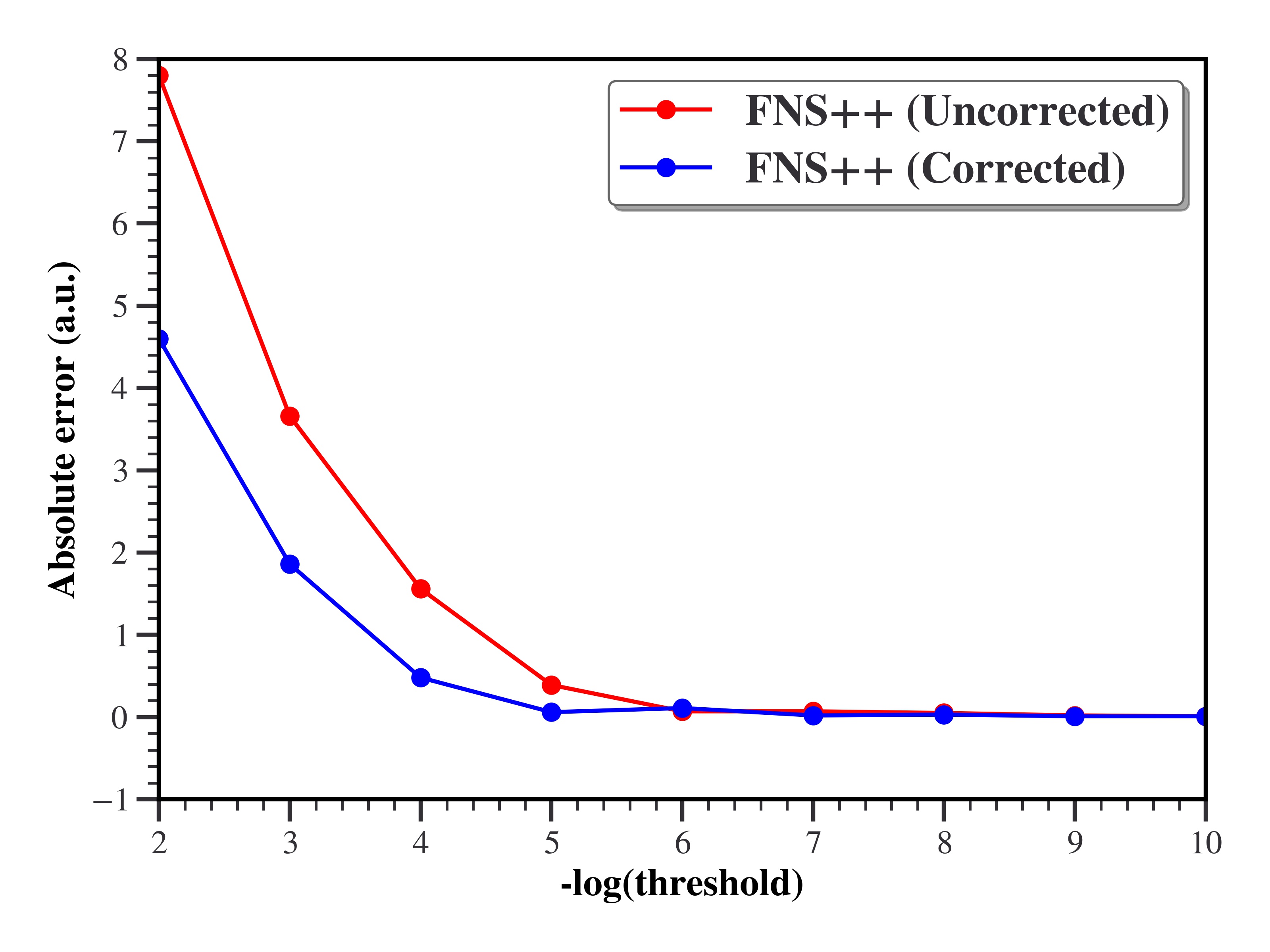} % Adjust width as needed
    \caption{\label{fig:epsart} Absolute error of dynamic polarizability at 0.072 a.u. for Zn atom at uncontracted aug-cc-pVDZ/4c-LRCCSD level in the FNS++ frameworks.}
\end{figure}

\subsection{Benchmarking basis sets}

The choice basis set is known to play an important role in determining the accuracy of the calculated polarizability values, and one needs to augment the basis function with a sufficient number of diffuse functions to obtain the desired accuracy\cite{LARSEN1998536}.  We have calculated the dynamic polarizability of the Kr atom at 10$^{-5}$ FNS++ occupation threshold using 4c-LRCCSD theory. The experimental value is 17.075 a.u., and Table I shows the variation in polarizability values of the Kr atom with the gradual increase in the basis set cardinal number and augmentation level. The accuracy of the calculated dynamic polarizability improves as the basis set size increases from double zeta to quadruple zeta level. Without the augmentation, the results significantly deviate from the experimental value, with absolute errors ranging from 9.77 a.u. to 2.306 a.u. when moving from double zeta to quadruple zeta. In the case of dyall.v2z, even the single augmentation leads to significant improvement of the polarizability values. The variation in the polarizability values with the cardinal number are smaller when the basis set is augmented. The results converge with doubly augmented dyall.v4z basis set.  So, we have chosen d-aug-dyall.v4z as the final basis set for all the subsequent calculations discussed in the next section, unless stated otherwise.
The inclusion of perturbative correction results in a shift of $\sim$ 0.1 a.u. for all the basis sets.

\begin{table}
\caption{\label{tab:table1}Basis set benchmarking for dynamic polarizability of Kr at 0.072 a.u. frequency using 4c-LRCCSD at 10$^{-5}$ FNS++ occupation threshold. The experimental value is 17.075 a.u.\cite{hohm1990interferometric}}
\begin{ruledtabular}
\begin{tabular}{ccccccc}
 & \multicolumn{2}{c}{dyall.v2z}& \multicolumn{2}{c}{dyall.v3z}& \multicolumn{2}{c}{dyall.v4z}\\ 
 &CC\footnotemark[1]&CC\footnotemark[2]&CC\footnotemark[1]&
 CC\footnotemark[2]&CC\footnotemark[1]&CC\footnotemark[2]\\
\hline
no-aug& 7.298& 7.323& 12.134&12.160& 14.769& 14.661\\ 
s-aug& 16.878& 16.733& 17.408&17.302& 17.302& 17.241\\  
d-aug& 17.328& 17.101& 17.551&17.429& 17.393& 17.261\\ 
t-aug& 17.330& 17.102& 17.552&17.436& 17.395& 17.265\\ 
q-aug& 17.270& 17.086& 17.553&17.437& 17.396& 17.265\\ 
\end{tabular}
\end{ruledtabular}
\footnotetext[1]{Uncorrected}
\footnotetext[2]{Corrected}
\end{table}

\subsection{Static and dynamic polarizabilities of Group IIB atoms in the FNS++ basis}

\begin{table}
\caption{\label{tab:table2} Static and dynamic polarizability (a.u.) of Zn, Cd, Hg using 4c-LRCCSD at 10$^{-5}$ FNS++ occupation threshold with d-aug-dyall.v4z basis set}
\begin{ruledtabular}
\begin{tabular}{cccccc} 
 &frequency (a.u.) &4c-CC\footnotemark[1]&4c-CC\footnotemark[2]& X2C-CC\footnotemark[3]&
 Expt.\\
\hline
Zn& 0.00000& 39.70& 39.41&40.42&38.8 $\pm$ 0.8\cite{goebel1996theoretical}\\ 
& 0.07198& 44.16& 43.78&&43.03 $\pm$ 0.32\cite{goebel1996theoretical}\\  
& 0.08383& 46.05& 45.65&&44.76$\pm$ 0.31\cite{goebel1996theoretical}\\ 
& 0.14014& 66.36& 65.79&&63.26$\pm$ 0.12\cite{goebel1996theoretical}\\ 
& & & &\\
 Cd& 0.00000& 47.21& 47.09&48.25&47.5 $\pm$ 2\cite{hohm2022dipole}\\
 & 0.07198& 52.97& 52.78&&54.2 $\pm$ 0.95\cite{goebel1995dispersion}\\
 & 0.08383& 55.45& 55.24&&56.23 $\pm$ 0.38\cite{goebel1995dispersion}\\
 & 0.14014& 81.39& 81.06&&68.8 $\pm$ 2.3\cite{goebel1995dispersion}\\
 & & & &\\
 Hg& 0.00000& 35.13& 34.99&35.25&33.92$\pm$ 0.34\cite{goebel1996dipole}\\
 & 0.07198& 37.37& 37.18&&35.75$\pm$ 0.310\cite{goebel1996dipole}\\
 & 0.08383& 38.26& 38.07&&36.63$\pm$ 0.317\cite{goebel1996dipole}\\
 & 0.14014& 46.69& 46.43&&44.64$\pm$ 0.331\cite{goebel1996dipole}\\ 
\end{tabular}
\end{ruledtabular}
\footnotetext[1]{Uncorrected}
\footnotetext[2]{Corrected,}
\footnotetext[3]{X2C Hamiltonian with s-aug-dyall.v2z basis set\cite{yuan2024formulation}}
\end{table}
Table II presents the static and dynamic polarizabilities of Zn, Cd, and Hg at 4c-LRCCSD level, calculated using the d-aug-dyall.v4z basis set and 10$^{-5}$ threshold for the FNS++ approach. Both corrected and uncorrected values are provided alongside experimental data for comparison. Three dynamic polarizabilities are computed at specific experimental frequencies, while the static polarizability corresponds to zero frequency. The X2C-CC values for the static polarizability are taken from the work of Gomes and co-workers.\cite{yuan2024formulation} For the Zn atom, both the uncorrected and corrected static polarizability values are in good agreement with the experimental data, demonstrating reasonable accuracy, with an overestimation of 0.896 a.u. and 0.612 a.u., respectively, which suggests that the correction may improve the accuracy of the calculated static polarizability. For the dynamic polarizability of the Zn atom, the values at frequencies of 0.07198 a.u., 0.08383 a.u., and 0.14014 a.u.\ also show good agreement with the experimental results. It can be seen that the magnitude of the perturbative correction increases for the dynamic polarizability with the increase in the frequency. For Cd, the uncorrected value of the static polarizability is 47.21 a.u., and the corrected value is 47.09 a.u., both in good agreement with the experimental value of 47.5 ± 2 a.u.. The dynamic polarizabilities also show reasonable agreement with the experiment. At \( \omega = 0.07198 \) a.u., the uncorrected value of 52.97 a.u.\ underestimates the experimental value by 1.23 a.u., and the corrected value is 52.78 a.u. At \( \omega = 0.08383 \) a.u., the uncorrected value of 55.45 a.u.\ underestimates the experimental result by 0.78 a.u., while the corrected value is 55.24 a.u. However, at  $\omega$ = 0.14014 a.u., both the uncorrected and corrected values (81.31 a.u.\ and 81.06 a.u., respectively) overestimate the experimental value (68.8 ± 2.3 a.u.) significantly, with deviations of 12.59 a.u. and 12.263 a.u., respectively. A significant deviation is observed in the dynamic polarizability of Cd at a frequency of 0.14014 a.u., which can be attributed to the presence of a pole near this frequency. In the case of Hg, the static polarizability values of 35.13 a.u. (uncorrected) and 34.99 a.u. (corrected) are in reasonable agreement with the experimental value of 33.92 a.u., with the corrected value being closer to the experiment. For the dynamic polarizabilities at \( \omega = 0.07198 \) a.u., the uncorrected value of 37.37 a.u. overestimates the experimental value by 1.62 a.u., and the corrected value of 37.18 a.u.\ brings the error down to 1.432 a.u. At \( \omega = 0.08383 \) a.u., the uncorrected value of 38.26 a.u.\ overestimates the experimental value by 1.634 a.u., while the corrected value of 38.069 a.u.\  still shows an overestimation, although the difference is slightly smaller at 1.43 a.u. At \( \omega = 0.14014 \) a.u., the uncorrected value of 46.69 a.u. overestimates the experimental value by 2.05 a.u., while the perturbative correction reduces the error to 1.79 a.u. The agreement between the calculated static and dynamic polarizability values with the experiment is better in the FNS++LR-CCSD than the X2C-LR-CCSD method\cite{yuan2024formulation}, presumably due to the use of a larger basis set in the present study.

\begin{table}
\caption{\label{tab:table1}Static polarizability (a.u.) of hydrogen halides at 10$^{-5}$ FNS++ occupation threshold using 4c-LRCCSD.}
\begin{ruledtabular}
\begin{tabular}{ccccc} 
 &CC\footnotemark[1]&CC\footnotemark[2]&
 CC\footnotemark[3]& Expt.\cite{hohm2013experimental}\\
\hline
HF& 5.05 & 5.77&5.69& 5.6 $\pm$ 0.10\\ 
HCl&16.09 &17.63&17.56& 17.39 $\pm$ 0.20  \\  
HBr&22.58 &24.06&23.96& 23.74 $\pm$ 0.50\\ 
HI& 34.30 &35.71&35.69& 35.3$\pm$ 0.50\\ 
\end{tabular}
\end{ruledtabular}
\footnotetext[1]{X2C Hamiltonian with s-aug-dyall.v2z basis set\cite{yuan2024formulation}}
\footnotetext[2]{FNS++4c-LRCCSD (Uncorrected), d-aug-dyall.v4z}
\footnotetext[3]{FNS++4c-LRCCSD (Corrected), d-aug-dyall.v4z}
\end{table}

\subsection{Static and Dynamic Polarizability of Molecules}

Next, we turn our attention to the static polarizabilities of hydrogen halides. In Table III, we provide the computed polarizabilities for a series of hydrogen halides using the d-aug-dyall.v4z basis set with an occupation threshold of \( 10^{-5} \). For comparison, we also include previously reported LRCCSD values calculated using the X2C Hamiltonian with the s-aug-dyall.v2z basis set,\cite{yuan2024formulation} as well as experimental results. It can be seen that FNS++ 4c-LRCCSD method gives accurate dynamic polarizability values, which are within the experimental error bar. For all four hydrogen halides, the X2C-LR-CCSD results show greater deviation from the experimental values, presumably due to the use of a smaller basis set. Figure 4 shows the percentage error in the static polarizability of hydrogen halides relative to the experimental values, computed using X2C and FNS++4c-LR-CCSD with the s-aug-dyall.v2z and d-aug-all.v4z basis sets, respectively.  It is evident that, at the occupation threshold of 10\(^{-5}\), the MP2-corrected results consistently exhibit significantly lower errors compared to the uncorrected values for all the molecules. The highest error reduction due to  MP2-level correction is observed for HF,  where the magnitude of reduction is approximately 1.42\% . When examining the polarizability anisotropies of hydrogen halides (see Table SI in the SI), it has been observed that all of them show good agreement with experimental values, except for HI. In the case of HI, anisotropy using d-aug-dyall.v4z and 4c-LRCCSD with the 10$^{-5}$ occupation threshold is 2.77 a.u, whereas the available experimental value is 11.4 a.u. A similar trend is also observed by Gomes and coworkers \cite{yuan2024formulation} in their X2C-LRCCSD implementation. The reason behind the discrepancy with the experimental results is not  clear at this moment.

\begin{figure}
\centering
    \includegraphics[width=0.5\textwidth]{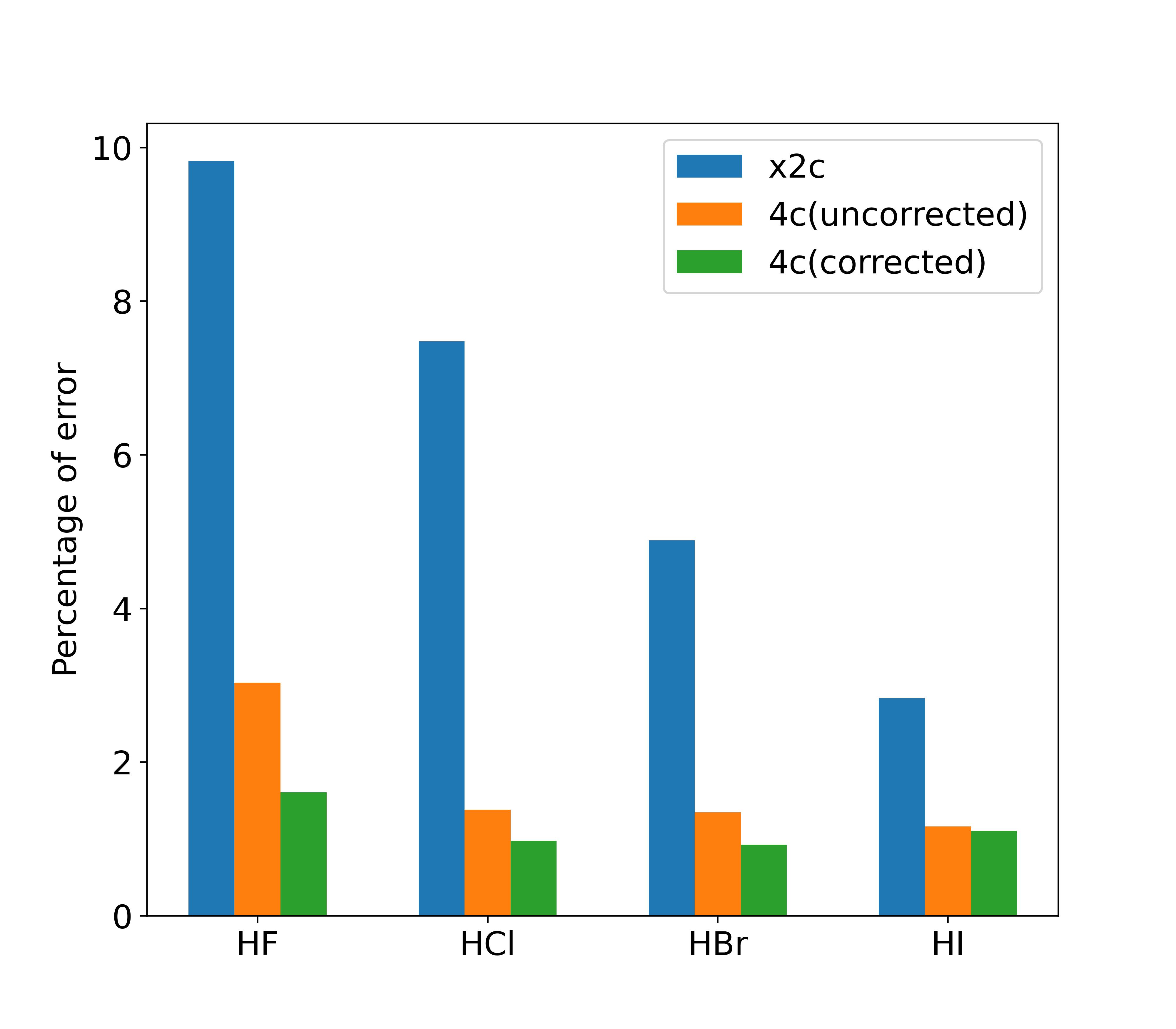} % Adjust width as needed
    \caption{\label{fig:epsart}Percentage error in static polarizability with respect to experiment for hydrogen halides using X2C(s-aug-dyall.v2z) and FNS++4c-LRCCSD (d-aug-dyall.v4z) with and without perturbative correction.}
\end{figure}

\begin{table}
\caption{\label{tab:table1}Static polarizability (a.u.) of dihalogens using10$^{-5}$ FNS++ occupation threshold at 4c-LRCCSD/d-aug-dyall.v4z level of theory. }
\begin{ruledtabular}
\begin{tabular}{cccccc} 
 &$\alpha_{\perp}$&$\alpha_{\parallel}$&$\alpha$ &\% error& Expt.\cite{hohm2013experimental}\\
\hline
F$_{2}$& 6.58& 12.52& 8.56&2.15&8.38 $\pm$ 0.15\\ 
Cl$_{2}$& 25.36& 42.59& 31.11&2.22& 30.43 $\pm$ 0.30\\  
Br$_{2}$& 36.32& 62.84& 45.16&4.93& 47.5 $\pm$ 1.1\\ 
I$_{2}$& 56.47& 102.22& 71.72&2.89&69.7 $\pm$ 1.8\\
ICl& 41.03& 67.01& 49.69&13.45&43.8 $\pm$ 4.4\\
\end{tabular}
\end{ruledtabular}
\end{table}

To further demonstrate the applicability of our implementation, representative calculations on halogen dimers are also performed.  Table IV provides the parallel ( \(\alpha_{\parallel}\)) and perpendicular (\(\alpha_{\perp}\)) components of the static polarizability and the isotropic polarizability (\(\alpha = (\alpha_{\perp} + 2\alpha_{\parallel})/3\)) for dihalogens (F\(_2\), Cl\(_2\), Br\(_2\), I\(_2\)) and the interhalogen ICl. The calculations were performed at the 4c-LRCCSD/d-aug-dyall.v4z level of theory with a \(10^{-5}\) occupation threshold. Experimental isotropic polarizabilities are also included in Table IV for comparison. The calculated isotropic polarizability (\(\alpha\)) follows the trend: F\(_2\) < Cl\(_2\) < Br\(_2\) < I\(_2\). This trend is consistent with the increasing size and electron count of the dihalogens as we move down Group 17 in the periodic table. Larger atoms with more diffuse electron clouds exhibit greater polarizability due to their increased ease of electron cloud distortion. The polarizability anisotropy increases significantly down the group, with F\(_2\) showing the smallest anisotropy (5.94 a.u.) and I\(_2\) the largest (45.75 a.u.). This behavior is indicative of the greater elongation of the electron cloud and polarizability along the molecular axis for heavier halogens, owing to their more extended bond lengths and greater atomic polarizabilities. ICl has an isotropic polarizability (\(\alpha = 49.69\)) between those of Cl\(_2\) and Br\(_2\), which is expected given its intermediate molecular size and composition. The anisotropy for ICl (25.98 a.u.) is larger than that of Cl\(_2\) but smaller than that of I\(_2\), consistent with its heteronuclear nature and bond characteristics. The deviations with respect to experiment for F\(_2\), Cl\(_2\), Br\(_2\), I\(_2\) are all within acceptable ranges (less than 5\%), confirming the accuracy of the current implementation. The slightly higher error for ICl may arise from the missing higher-order correlation effects or missing vibrational effects that were not considered in the present study. However, it should be noted that the experimental error bar for the ICl is also high in comparison to the other dihalogens.

\begin{table}
\caption{\label{tab:table1}Dynamic polarizability (a.u.) of I$_{2}$ using 10$^{-5}$ FNS++ occupation threshold at 4c-LRCCSD/d-aug-dyall.v4z level }
\begin{ruledtabular}
\begin{tabular}{cccc} 
 &$\alpha_{\perp}$&$\alpha_{\parallel}$&$\alpha$\\
 \hline
 \multicolumn{4}{c}{frequency = 0.07198 a.u.}\\
\hline
X2C-HF\footnotemark[1]& 55.0& 152.0& 87.4\\ 
X2C-B3LYP\footnotemark[1]& 58.7& -10.7& 35.6\\  
X2C-CC\footnotemark[1]& 55.8& 114.8& 75.5\\ 
4c-CC\footnotemark[3]& 57.2& 114.18& 76.19\\
 Expt.\cite{maroulis1997electrooptical}& & &86.8 $\pm$ 2.2\\
 \hline
 \multicolumn{4}{c}{frequency = 0.07669 a.u.}\\
\hline
 X2C-HF\footnotemark[1]& 56.0& -97.3&4.9\\
 X2C-B3LYP\footnotemark[1]& 62.0& 75.4&66.5\\
 X2C-CC\footnotemark[1]& 56.8& 124.0&79.2\\
 4c-CC\footnotemark[3]& 58.4& 119.1&78.7\\
 Expt.\cite{maroulis1997electrooptical}& & &93.6 $\pm$ 3.4\\
 \hline
 \multicolumn{4}{c}{frequency = 0.14014 a.u.}\\
\hline
 X2C-HF\footnotemark[1]& 55.3& 117.9&76.2\\
 X2C-B3LYP\footnotemark[1]& 61.0& 114.5&78.8\\
 X2C-CC\footnotemark[1]& 59.9& 113.9&77.9\\
 4c-CC\footnotemark[3]& 63.0& 117.6&81.2\\
 Expt.\cite{maroulis1997electrooptical}& & &95.3 $\pm$ 1.9\\
\end{tabular}
\end{ruledtabular}
\footnotetext[1]{Using X2C Hamiltonian. Taken from ref \cite{yuan2024formulation}}
\footnotetext[2]{Using 4c-DC Hamiltonian at 10$^{-5}$ FNS++ occupation threshold (Uncorrected)}
\end{table}

Table V presents the perpendicular (\(\alpha_{\perp}\) ) and parallel (\(\alpha_{\parallel}\)) component of dynamic polarizability components and the isotropic polarizability (\(\alpha\)) of I\(_2\) at three different frequencies: 0.07198 a.u., 0.07669 a.u., and 0.14014 a.u.. The calculations are performed at the 4c-LRCCSD/d-aug-dyall.v4z level of theory with a 10$^{-5}$ FNS++ occupation threshold. At a frequency of 0.07198 a.u., the isotropic polarizabilities calculated using the CC-based methods with 4c and X2C Hamiltonians are generally consistent with each other. These values are slightly lower than the experimental value of \(86.8 \pm 2.2\) a.u., and the 4c-LRCCSD result is  closer to the experiment compared to the X2C-LRCCSD level of calculation due to the larger basis set used in the former. Notably, the B3LYP result significantly deviates from the experiment, yielding an isotropic polarizability of 35.6 a.u.. This discrepancy highlights the limitations of density functional methods, particularly for systems where both relativistic and electron correlation effects play a significant role. As the frequency increases to 0.07669 a.u., the isotropic polarizability values calculated using CC methods show an increase. The range of CC values remains consistent across different Hamiltonians, although these values still fall below the experimental result of \(93.6 \pm 3.4\) a.u. Interestingly, the B3LYP result improves significantly at this frequency, yielding an isotropic polarizability of 66.5 a.u., though it remains much lower than the experimental value. The HF method shows large deviations, with an isotropic polarizability of only 4.9 a.u., driven by a negative \(\alpha_{\parallel}\) component. These results highlight the importance of electron correlation, which is not captured by HF and is only partially accounted for by B3LYP. At the highest frequency considered (0.14014 a.u.), the deviation in the isotropic polarizability values obtained from various CC methods are slightly higher, with the 4c-LRCCSD result (\(81.2\) a.u.) being the closest to the experimental value of \(95.3 \pm 1.9\) a.u.. The B3LYP method yields an isotropic polarizability of 78.8 a.u., showing better agreement with the CC results but still underestimating with respect to the experimental value. Across all frequencies, the calculated \(\alpha_{\parallel}\) components are systematically larger than \(\alpha_{\perp}\), reflecting the anisotropic nature of the polarizability tensor in I\(_2\). This anisotropy is more pronounced at lower frequencies, where the difference between \(\alpha_{\parallel}\) and \(\alpha_{\perp}\) is the largest, particularly in the X2C-LRCCSD and 4c-LRCCSD results. The experimental isotropic polarizabilities are consistently higher than the theoretical values, particularly at higher frequencies, suggesting the need for further improvements in the protocol for simulation with the inclusion of higher-order excitation in the coupled cluster calculations (e.g., CCSDT or CC3) and vibrational corrections. 
It is interesting to note that with the default threshold settings, the number of virtual spinors included in the correlation space was restricted to 248, 250, and 254 for the calculations at frequencies of 0.07198 a.u., 0.07669 a.u., and 0.14014 a.u., respectively. This corresponds to approximately 25\% of the total 1002 virtual spinors being included in the correlation. Despite this significant reduction, the present implementation produces results that are both reasonably accurate and consistent with previous findings, highlighting the efficiency and robustness of the current approach.

\begin{table}
\caption{\label{tab:table1}Static polarizability (a.u.) of AuH, AuF, AuCl, and HgCl$_{2}$ using 10$^{-5}$ FNS++ occupation threshold at 4c-LRCCSD/d-aug-dyall.v4z level.}
\begin{ruledtabular}
\begin{tabular}{ccccc} 
 &$\alpha_{\perp}$&$\alpha_{\parallel}$&$\alpha$& Expt.\\
\hline
AuH& 36.94& 41.23& 38.37&-\\ 
AuF& 30.57& 39.05& 33.40&-\\  
AuCl& 41.06& 64.94& 49.02&-\\ 
HgCl$_{2}$& 46.39& 98.06& 63.61&61.2 $\pm$ 1.3\cite{hohm2013experimental}\\
\end{tabular}
\end{ruledtabular}
\footnotetext[1]{Uncorrected}
\end{table}

Next, we have tested our implementation with molecules containing 3d metals, such as Au and Hg. Table \ref{tab:table1} presents the calculated static polarizability components (\(\alpha_{\perp}\) and \(\alpha_{\parallel}\)) and the isotropic polarizability of AuH, AuF, AuCl, and HgCl\(_2\) at the 4c-LRCCSD/d-aug-dyall.v4z level using a 10\(^{-5}\) occupation threshold. The experimental isotropic polarizability value is available only for HgCl\(_2\), enabling a direct comparison with the computed result. The isotropic polarizability values exhibit systematic trends across the molecules. Among the molecules containing Au, AuF has the lowest isotropic polarizability (\(33.4\) a.u.), reflecting its compact nature and the relatively small extent of its electron cloud distortion. The smaller \(\alpha_{\perp}\) and overall polarizability in AuF can be attributed to the strong electronegativity of fluorine, which leads to tighter binding of electrons and reduced polarizability. For AuCl, a significant increase in polarizability is observed (\(\alpha = 49.02\) a.u.), consistent with the larger atomic size and less electronegative nature of chlorine compared to fluorine. The large anisotropy (\(\Delta \alpha = 23.88\) a.u.) indicates a pronounced directional dependence of the polarizability in this molecule. No experimental results are available for AuH, AuF, and AuCl.  However,  the HgCl\(_2\), which exhibits the highest isotropic polarizability (\(63.62\) a.u.) among the studied molecules, shows excellent agreement with the experimental value of \(61.2 \pm 1.3\) a.u., with a deviation of approximately 4\%. 
\begin{figure}
\centering
    % First subplot
    \begin{subfigure}{0.48\textwidth} % Adjust width as needed
        \includegraphics[width=\linewidth]{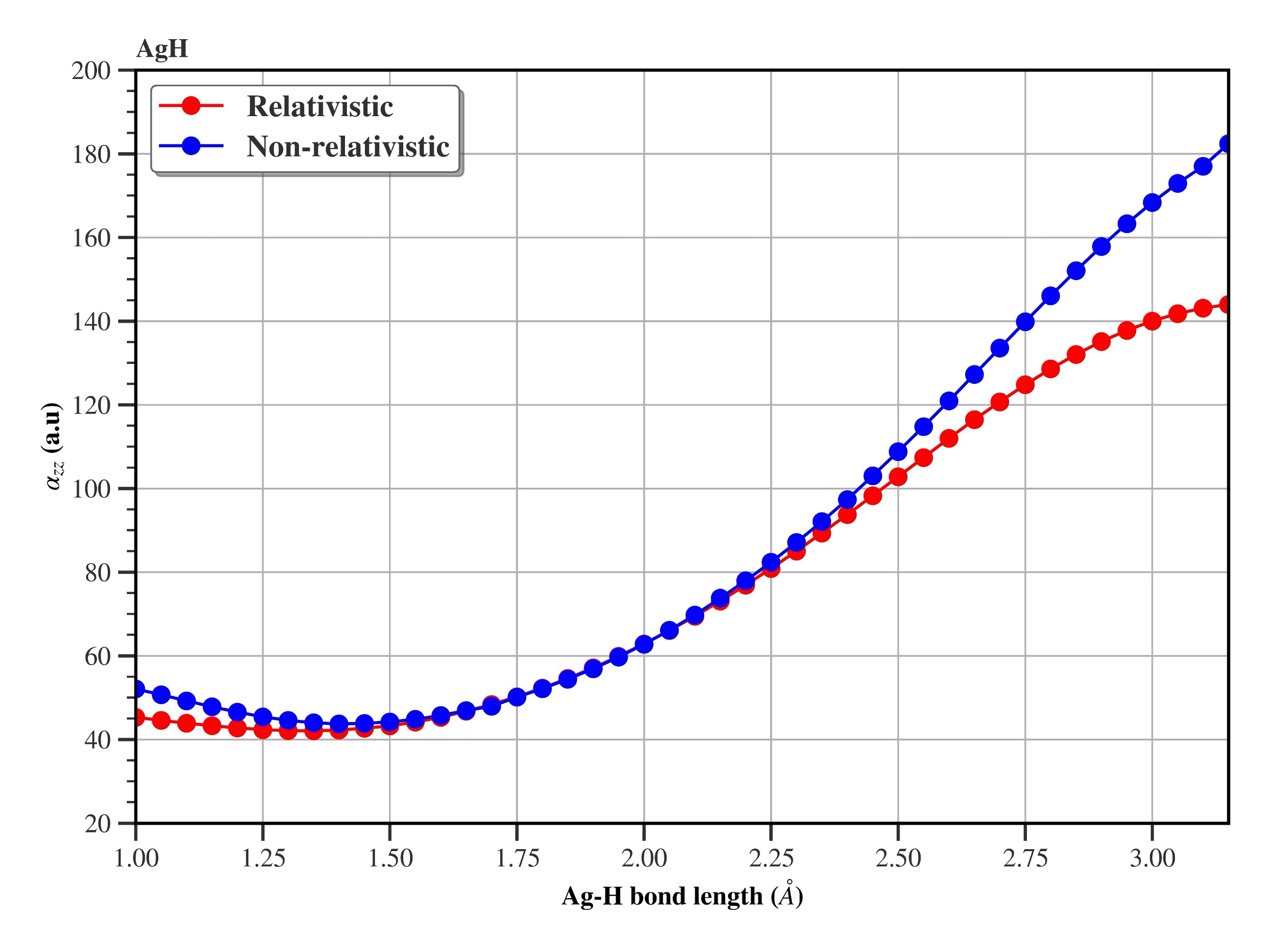} % Replace with your file
        \caption{}
        \label{fig:sub1}
    \end{subfigure}
    \begin{subfigure}{0.48\textwidth}
        \includegraphics[width=\linewidth]{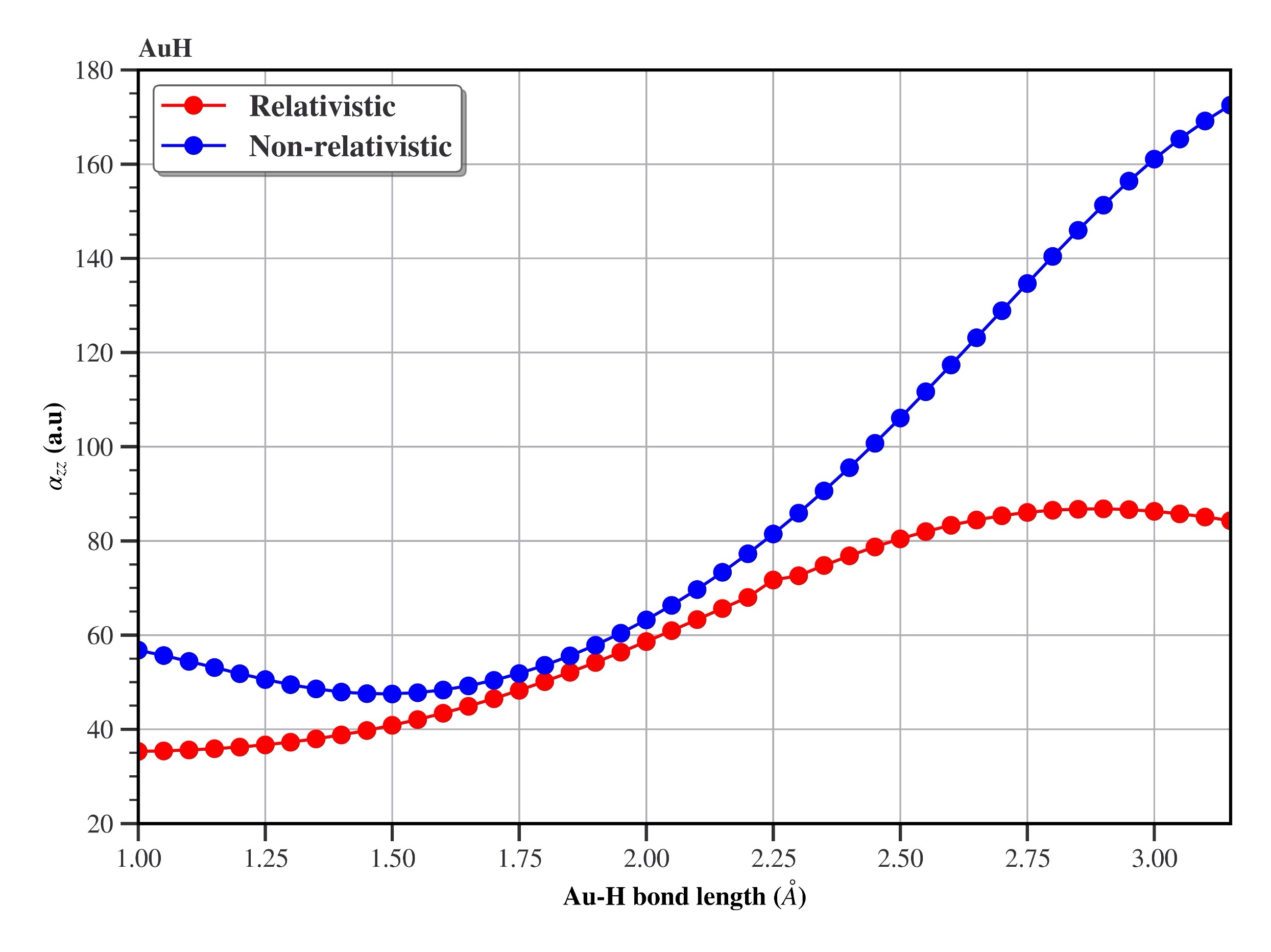}
        \caption{}
        \label{fig:sub2}
    \end{subfigure}

    % Main caption for the figure
    \caption{\label{fig:epsart}Static Polarizability spectrum of the parallel component of (a)AgH and (a)AuH with bond length at s-aug-dyall.v2z/4c-LRCCSD level with 10$^{-5}$ FNS++ occupation threshold.}
\end{figure}
Figure 5 illustrates the static polarizability of the parallel component for AgH and AuH molecules as a function of bond length, computed using both relativistic and non-relativistic methods at the s-aug-dyall.v2z/4c-LRCCSD level with a 10$^{-5}$ FNS++ occupation threshold. These results reveal significant differences between the two approaches, particularly highlighting the importance of relativistic effects in systems involving heavy atoms. For AgH, the polarizability increases with bond length in both relativistic and non-relativistic cases. However, the relativistic polarizabilities are consistently lower, with the difference between the two methods becoming more pronounced at larger bond lengths. This suggests that relativistic effects temper the polarizability by reducing the electronic cloud's response to external fields, particularly as the bond elongates. In contrast, the non-relativistic polarizabilities exhibit a steeper rise, leading to an overestimation of the molecular response. In AuH, the trends are qualitatively similar but more pronounced due to the heavier gold nucleus, which intensifies relativistic effects such as spin-orbit coupling and orbital contraction. While both approaches show an increase in polarizability with bond length, the relativistic values for AuH plateau beyond a bond length of approximately 2.00 Å. This saturation is absent in the non-relativistic calculations, which continue to overestimate the polarizability. The plateau observed in the relativistic case reflects the stabilization and contraction of the $d$ and $f$ orbitals, a phenomenon intrinsic to gold under relativistic conditions. These results underscore the necessity of incorporating relativistic corrections for accurate modeling of polarizabilities in heavy-element containing systems. Non-relativistic methods could lead to incorrect predictions in molecular response properties.

\subsection{Simulation of Optical Trapping of Molecules}

In recent years, there has been a remarkable advancement in the field of atomic quantum gases, and a key objective has been to attain control over individual quantum degrees of freedom in molecular systems. Owing to the exceptional control over these degrees of freedom, ultracold molecules in optical lattices opens up diverse and exciting avenues for exploration in modern chemistry and physics
,\cite{carr2009focus, krems2008cold} which includes the creation of molecular Bose-Einstein condensates,\cite{jaksch2002creation} high-precision measurements of fundamental constants,\cite{chin2009ultracold, zelevinsky2008precision, demille2008enhanced} and applications in quantum computing.\cite{demille2002quantum,yelin2006schemes} Optically trapped molecules in the vibrational ground state in the lowest singlet or triplet state have been successfully prepared by a number of research groups in recent years\cite{ni2008high,danzl2010ultracold,lang2008ultracold, takekoshi2014ultracold,molony2014creation}. For these types of ultracold experiments, dipole traps or optical lattices are used.\cite{yan2013observation,zhu2014suppressing}
In such cases, accurate knowledge of the dynamic polarizability is crucial for ensuring precise control over experimental conditions. We have employed our 4c-LRCCSD implementation to estimate the optimal parameter for trapping of Rb dimers in an optical lattice that ensures equal trapping of both the initial and final molecular states involved in the stimulated Raman adiabatic passage (STIRAP) process.\cite{bergmann1998coherent} At a particular wavelength --- the ``magic wavelength'' --- the ac Stark shift of the dimer’s final state becomes identical to that of the initial state consists of a pair of unperturbed atoms. Feshbach and ground-state molecules must be trapped in a 3D optical lattice with identical dynamic polarizabilities at the lattice wavelength to provide optimal control. 

Here we have calculated the dynamic polarizabilities of the Rb$_{2}$ molecule and the Rb atom using 4c-LRCCSD and s-aug-dyall.v2z basis set with 10$^{-5}$ FNS++ occupation threshold. Figure 7 depicts the dynamic polarizability of Rb$_{2}$ together with the dynamic polarizability of a Rb-atom pair (2$\alpha_{Rb}$) simulating a Feshbach molecule within a frequency range of 0.002-0.065 a.u.. From the figure, it is evident that the dynamic polarizabilities of both systems are in similar magnitudes in the low-frequency range and they start to differ when the frequency reaches the resonant region. We have observed that the polarizability of both species is 1984 a.u.\ at a frequency of 0.04674 a.u. This frequency, corresponding to the ``magic wavelength,'' is where the polarizability of the ground-state molecules matches that of a Feshbach molecule. In previous experimental studies,\cite{danzl2010ultracold, markus2015} they have used 0.0428 a.u.\  as the frequency of the optical lattice and our calculated magic wavelength using 4c-LRCCSD using FNS++ basis is closer to this value, indicating that the Rb$_{2}$ molecule can be a good candidate for this kind of ultracold experiment. The small discrepancy of approximately 0.00394 a.u.\  between the computed magic wavelength and the experimental trap laser frequency may stem from factors such as basis set size, and vibrational effects.

\begin{figure}
\centering
    \includegraphics[width=0.5\textwidth]{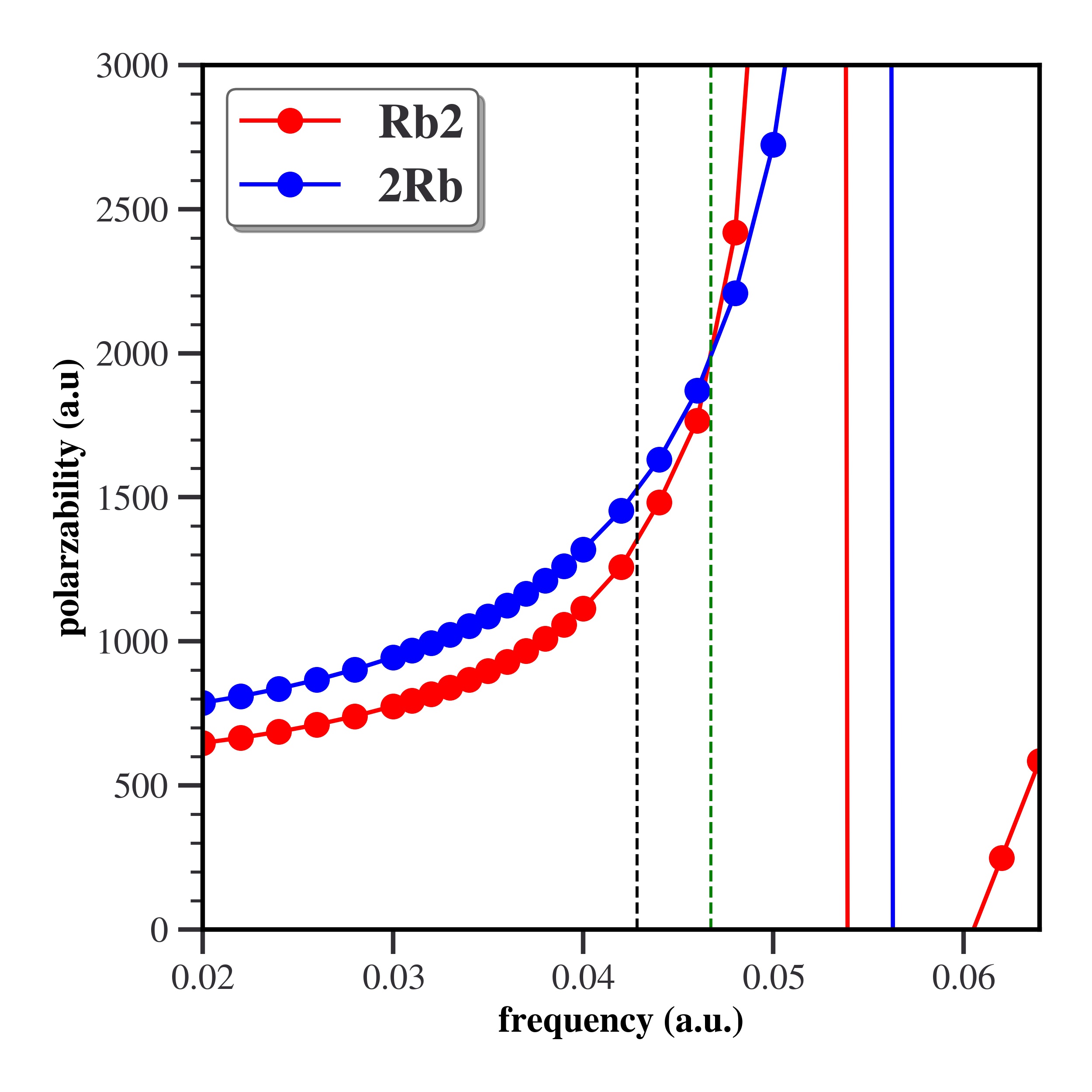} % Adjust width as needed
    \caption{\label{fig:epsart}Polarizability spectrum of Rb$_{2}$ molecule (red line) and Feshbach molecule corresponding to 2$\alpha_{Rb}$ (blue line) using s-aug-dyall.v2z with 10$^{-5}$ FNS++ occupation threshold using 4c-LRCCSD.}
\end{figure}

\subsection{Computational Efficiency}
One of the primary motivations for using the  FNS++ method is to reduce the computational cost of the polarizability calculations in large basis sets.  Figure 6 compares computational timings between the various steps of canonical 4c-LRCCSD and the FNS++ 4c-LRCCSD approach for HI in s-aug-dyall.v2z with a 10$^{-5}$ FNS++ occupation threshold. All the calculations were performed on a dedicated workstation with two Intel(R) Xeon(R) CPU Silver 4210R @ 2.40GHz CPU and 1000 GB of total RAM.  In this computational setting the canonical 4c-LRCCSD calculation involves a virtual space of 228 spinors and in the FNS++ basis it reduces to 60 only after truncation with 10$^{-5}$ occupation threshold.

The most noticeable improvement is observed in the response-function step. The canonical approach for this step requires nearly 1hr, 6min, 40 secs, while the FNS++ method achieves a significant reduction, completing the same step in just 36 secs. This striking reduction highlights the efficiency of virtual space truncation in the FNS++ framework for response property calculations. The CCSD and left-CCSD steps mentioned in Figure 6 imply the time required for solving ground state unperturbed $\hat{T}$ and $\hat{\Lambda}$ amplitude equations, which also reduced significantly in the FNS++ basis. Similarly, the integral transformation step, while still computationally demanding, shows a reduction of more than 50\% with FNS++, further emphasizing the method's efficiency. The integral transformation step in the canonical basis took approximately 20 minutes and 45 seconds, whereas in the FNS++ basis, it required only 8 minutes and 57 seconds. This represents an almost 2.3 times reduction in computational time for the integral transformation step.  The total time for the correlation calculation in the canonical 4c-LR-CCSD method requires 2 hrs, 20 mins, and 42 secs. The same calculation in the FNS++ basis only takes  13 mins, 4 secs, showcasing a factor of 10.8 speed up.

\begin{figure}
\centering
    % First subplot
    
        \includegraphics[width=\linewidth]{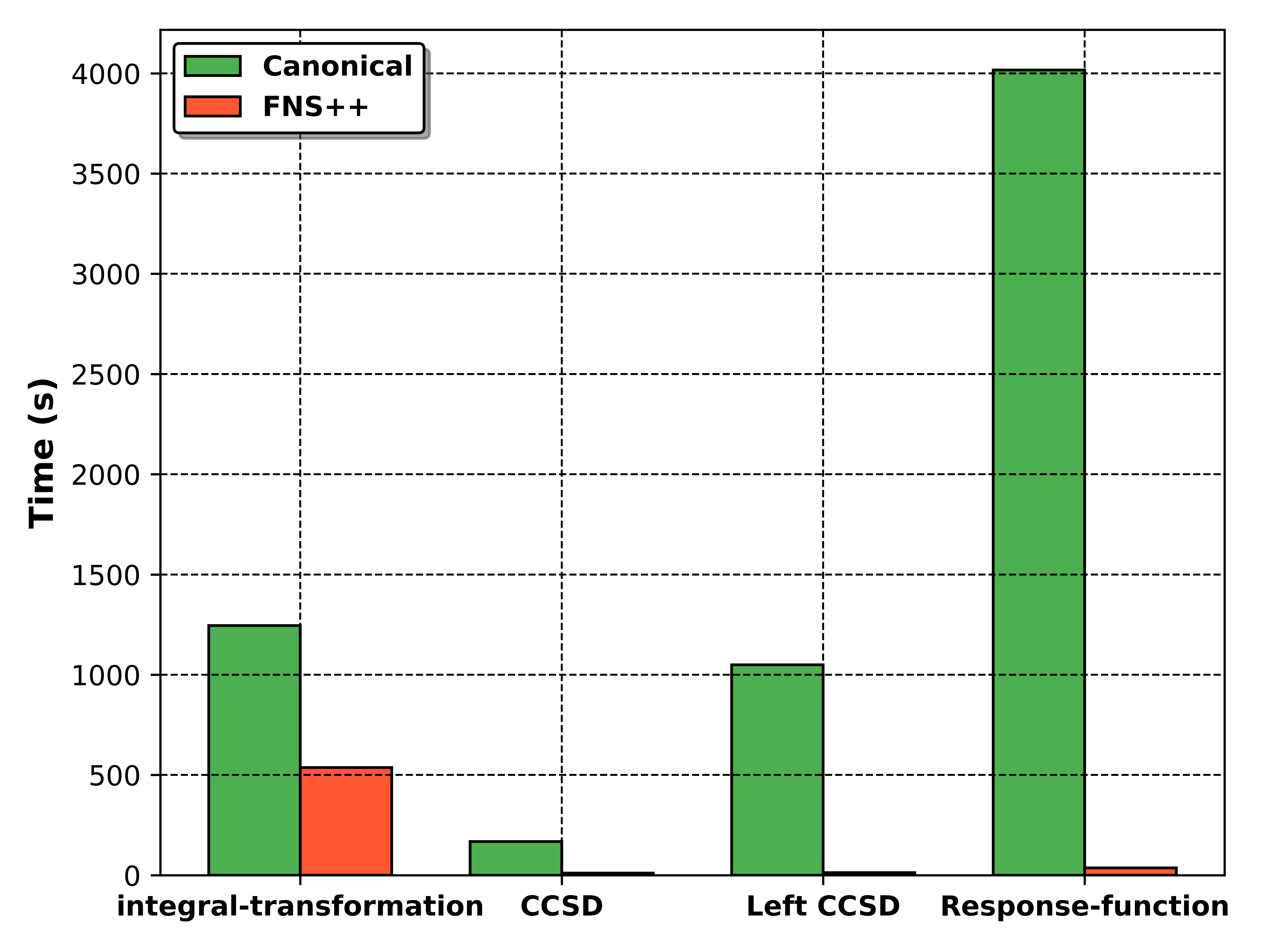}

    %\begin{subfigure}{0.47\textwidth}
    %    \includegraphics[width=\linewidth]{molecules_time.png}
    %    \caption{}
     %   \label{fig:sub2}
    %\end{subfigure}

    % Main caption for the figure
    \caption{\label{fig:epsart} Time taken for the individual computational steps for the static polarizability calculation of HI molecule using the s-aug-dyall.v2z basis set. CCSD and left-CCSD imply the time for solving ground state unperturbed $\hat{T}$ and $\hat{\Lambda}$ amplitude equations.}
\end{figure}

\section{Conclusion}
We present the theory, implementation, and benchmarking of a low-cost 4c-LRCCSD for the calculation of polarizability using a perturbation-sensitive natural spinor basis. We have investigated the optimal truncation threshold for linear response properties within the 4c-relativistic framework. It has been found that although the FNS-based scheme works very well for relativistic CCSD energy calculations, the ground state density constructed from the relativistic MP2 doubles amplitudes in the FNS scheme are not suitable for static or dynamic polarizability calculations. In the FNS-based truncation of the virtual space, highly diffuse virtual natural spinors with very low occupation numbers are removed. However, these spinors are crucial for accurately describing polarizability, leading to a larger error in the calculated values. In the current implementation, we have developed a perturbed ground-state MP2-level density, and the corresponding natural orbitals (called FNS++) are more suitable in capturing the wave function's response to the external perturbation.
Our 4c-LRCCSD implementation within the FNS++ framework employs a systematic approach to truncate the virtual space, ensuring computational efficiency with systematically controllable accuracy for static and dynamic polarizability calculations. It has been observed that while the FNS-based truncation does not give a systematic convergence for polarizability calculations and requires approximately 70-80\% of the virtual spinor space to be included in the correlation for accurate results. The FNS++ method, on the other hand, gives similar accuracy with only 30-40\% of the virtual spinor space. An FNS++ occupation threshold of \( 10^{-5} \) has proven to be an optimal balance between computational efficiency and accuracy, offering significant time savings without compromising the quality of the results. An MP2-level correction can improve the accuracy of the FNS++-4C-LRCCSD polarizability values, However, the corrections are not as significant as that observed for the correlation energy,  The performance of the FNS++-based 4c-LRCCSD method has been evaluated for static and dynamic polarizability of various atoms and molecules containing heavy elements, demonstrating excellent agreement with both experimental data and existing theoretical results. The comparison of computational times between the canonical 4c-LRCCSD and FNS++ 4c-LRCCSD methods highlights the significant efficiency improvements provided by the FNS++ framework, with a 10 fold reduction in computational time for a 4c-LRCCSD calculation of the HI molecule. This demonstrates the improved efficiency and reliability of the FNS++ approach in reducing the computational cost while maintaining accuracy.  Quantitative agreement with the experiment will require the inclusion of triples in the relativistic LR-CCSD calculations. Work is in progress towards that direction.

\begin{acknowledgments}
AKD, SC, and AM acknowledge the support from IIT Bombay, IIT Bombay Seed Grant (Project No. R.D./0517-IRCCSH0-040), CRG (Project No. CRG/2022/005672) and MATRICS (Project No. MTR/2021/000420) project of DST-SERB, CSIR-India (Project No. 01(3035)/21/EMR-II), DST-Inspire Faculty Fellowship (Project No. DST/INSPIRE/04/2017/001730), and ISRO for financial support, IIT Bombay super computational facility, and C-DAC. Supercomputing resources (Param Smriti and Param Bramha) for computational time.
AKD acknowledges the research fellowship funded by the EU NextGenerationEU through the Recovery and Resilience Plan for Slovakia under the project No. 09I03-03-V04-00117.
SC acknowledge Prime Minister's Research Fellowship (PMRF).
AM acknowledge CSIR SRF Fellowship.  TDC work supported by the U.S. National Science Foundation via
grant CHE-2154753.
\end{acknowledgments}

%\nocite{*}
\bibliography{aipsamp}% Produces the bibliography via BibTeX.

\end{document}

\title{\textbf{Supplementary Material: A low-cost four-component relativistic coupled cluster linear response theory based on perturbation sensitive natural spinors}}

\author{Sudipta Chakraborty}
\author{Amrita Manna}
\affiliation{%
Department of Chemistry, Indian Institute of Technology Bombay, Mumbai 400076, India%
}

\author{T. Daniel Crawford}
\affiliation{%
Department of Chemistry, Virginia Tech, Blacksburg, Virginia 24061, USA%
}

\author{Achintya Kumar Dutta}
\affiliation{%
Department of Chemistry, Indian Institute of Technology Bombay, Mumbai 400076, India%
}
\email{achintya@chem.iitb.ac.in}

\affiliation{%
Department of Inorganic Chemistry, Faculty of Natural Sciences, Comenius University Bratislava, 
Ilkovičova 6, Mlynská dolina 842 15 Bratislava, Slovakia%
}
\email{achintya@chem.iitb.ac.in}
\email{achintya.kumar.dutta@uniba.sk}

%\date{\today}

\maketitle

\clearpage

\begin{table}
\caption{\label{tab:table1}Anisotropic dipole polarizability (a.u.) of hydrogen halides at 10$^{-5}$ FNS++ occupation threshold using 4c-LRCCSD.}
\begin{ruledtabular}
\begin{tabular}{ccccc} 
 &X2C-CC\footnotemark[1]&4c-CC\footnotemark[2]&
 4c-CC\footnotemark[3]& Expt.\\
\hline
HF& 1.96 & 1.18&1.17& 1.62\cite{muenter1972polarizability}\\ 
HCl&2.38 &1.68&1.66& 2.10\cite{bridge1966polarization}\\  
HBr&2.30 &2.14&2.08& 1.7\cite{pinkham2008extracting}\\ 
HI& 2.51 &2.78&2.77& 11.4\cite{denbigh1940polarisabilities}\\ 
\end{tabular}
\end{ruledtabular}
\footnotetext[1]{X2C Hamiltonian with s-aug-dyall.v2z basis set\cite{yuan2024formulation}}
\footnotetext[2]{FNS++4c-LRCCSD (Uncorrected), d-aug-dyall.v4z}
\footnotetext[3]{FNS++4c-LRCCSD (Corrected), d-aug-dyall.v4z}
\end{table}

\section{References}
\bibliographystyle{aipnum4-1}  % JCP uses AIP style
\bibliography{references}  % Ensure you have a references.bib file